\documentclass[galaxies,review,accept,moreauthors]{Definitions/mdpi} 

\firstpage{1} 
\pubvolume{1}
\issuenum{1}
\articlenumber{0}
\pubyear{2024}
\datereceived{ } 
\daterevised{ } 
\dateaccepted{ } 
\datepublished{ } 
\hreflink{https://doi.org/} 

\Title{X-RAY POLARIZATION FROM MAGNETAR SOURCES}

\TitleCitation{Title}

\Author{Roberto Taverna$^{1}$\orcidA{} and Roberto Turolla$^{1, 2}$\orcidB{}}

\AuthorNames{Firstname Lastname, Firstname Lastname and Firstname Lastname}

\AuthorCitation{Lastname, F.; Lastname, F.; Lastname, F.}

\address{%
$^{1}$ \quad Department of Physics and Astronomy, University of Padova; taverna@pd.infn.it\\
$^{2}$ \quad Mullard Space Science Laboratory, University College London; turolla@pd.infn.it}

\corres{Correspondence: roberto.taverna@unipd.it}

\abstract{The launch of {\it IXPE} telescope in late 2021 finally  made  polarization measurements in the $2$--$8\,\mathrm{keV}$ band a reality, more than 40 years after the pioneering observations of the {\it OSO-8} satellite. In the first two years of operations {\it IXPE} targeted more than 60 sources, including four magnetars, neutron stars with magnetic fields in the petaGauss range. In this paper we summarize {\it IXPE} main findings and discuss their implications for the physics of ultra-magnetized neutron stars. Polarimetric observations confirmed theoretical predictions according to which X-ray radiation from magnetar sources is highly polarized, up to $\approx 80\%$, the highest value detected so far. This provides an independent confirmation that magnetars are indeed endowed with a super-strong magnetic field and that the twisted magnetosphere scenario is the most likely explanation for their soft X-ray emission. Polarization measurements allowed us to probe the physical conditions of the star's outermost layers, showing that the cooler surface regions are in a condensed state, with no atmosphere on top. Although no smoking-gun of vacuum QED effects was found, the phase-dependent behaviour of the polarization angle strongly hints that vacuum birefringence is indeed at work in magnetar magnetospheres.}


\begin{document}




\section{Introduction}

Magnetars are a peculiar class of neutron stars (NSs), discovered at the end of the 1970s and initially classified into two separate groups, the anomalous X-ray pulsars (AXPs) and the soft-gamma repeaters (SGRs) \cite{mer08,tzw15,kb17}. The former exhibit a behavior substantially similar to conventional, isolated X-ray pulsars \cite{burd+00,mer11,wang+20}, but their (persistent) X-ray luminosity exceeds the rotational energy loss rate (and for this they were dubbed ``anomalous''). 
The latter, instead, are characterized by the repeated emission of energetic, short bursts, peaked in the hard X-, soft gamma-rays, in some cases modulated at a precise frequency, which led to the association of these sources to NSs and set them apart from short gamma-ray bursts (GRBs), with which they have been initially associated. 

AXPs and SGRs shine in the soft X-rays ($0.5$--$10\,\mathrm{keV}$), with persistent luminosities $L_\mathrm X\sim 10^{33}$--$10^{36}\,\mathrm{erg\,s}^{-1}$. Their spectra are well fitted by the superposition of two components: a thermal one, represented by a blackbody (BB) with temperature $\sim 0.5$--$1\,\mathrm{keV}$ and a power-law (PL) tail, with photon index $\approx 2$--$4$ \cite{ok14}\footnote{See the on line magnetar catalog: \url{https://www.physics.mcgill.ca/\~pulsar/magnetar/main.html}}. Both types of source share a peculiar bursting activity, with the occurrence of three different types of phenomena, according to their duration and amount of released energy. Short bursts are the most frequent events, common to both AXPs and SGRs, with typical duration $\approx 0.1$--$1\,\mathrm{s}$ and luminosity $10^{38}$--$10^{41}\,\mathrm{erg\,s}^{-1}$. SGRs also exhibit longer and most powerful intermediate flares, with duration $\approx 1$--$10\,\mathrm{s}$ and luminosity $\approx 10^{41}$--$10^{43}\,\mathrm{erg\,s}^{-1}$, which often occur concurrently with a substantial number of short bursts in the so-called burst forests \cite{isr+08}. Finally, three SGRs emitted a giant flare, characterized by a strong initial spike, with peak luminosity up to $10^{47}\,\mathrm{erg\,s}^{-1}$, followed by a long pulsating tail, modulated at the spin frequency of the star, for a total duration of $\approx 100\,\mathrm{s}$ \cite{maz+79,hurl+99,palm+05}. 

\citet[\citealp{td95}]{dt92} first suggested that the phenomenology of these sources can be explained in terms of their ultra-strong magnetic fields \cite[hence the name ``magnetars''; alternative scenarios have been suggested, e.g. invoking the presence of quark stars,][]{iwa05,ouy07}. Indeed, magnetar spin periods ($P\approx 1$--$12\,\mathrm{s}$) and spin-down rates ($\dot{P}\approx 10^{-14}$--$10^{-10}\,\mathrm{s\,s^{-1}}$) point to magnetic field strengths of $\approx 10^{14}$--$10^{15}\,\mathrm{G}$ (for the dipole component), $\approx 100$-$1000$ times stronger than those of common radio pulsars. However, the magnetic field inside the star is expected to be even stronger (up to $10^{16}\,\mathrm{G}$), with a toroidal component at least of the same order of the poloidal one \cite{td01,brait09}. Such a strong magnetic field can deform the NS crust \cite{pp11}, transferring magnetic helicity to the external field. As a consequence, the star's magnetosphere becomes twisted and currents must flow along the closed field lines, in order to sustain the twist \cite{tlk02}. Thermal photons emitted from the cooling star surface encounter an optically thick medium for resonant Compton scattering (RCS) and repeated scatterings onto the magnetospheric charged particles can indeed produce the power-law tails observed in the soft X-ray persistent spectra of magnetars. Actually, charge carriers are most likely electron-positron pairs produced in the magnetosphere itself by one-photon pair production in the strong magnetic field \cite{bt07}; however, also the simplified scenario in which only electrons and ions lifted from the surface are present can well reproduce the spectral observations \cite{ntz08,fd11,tav+14}.

Although the spectral analysis allows one to constrain the magnetospheric parameters, their determination is intrinsically degenerate, especially concerning the geometry of the system. A way out comes from polarization measurements. Photons propagating in a strongly magnetized environment are expected to be linearly polarized in two normal modes, the ordinary (O) and the extraordinary (X) one, parallel or perpendicular to the plane of the local magnetic field and the photon propagation direction, respectively \cite{gp74}. In a strong magnetic field, the cross sections of X-mode photons are strongly reduced with respect to those of the ordinary ones below the electron cyclotron frequency \cite{her79,vent79,mesz92,ntz08}. Moreover, close to the surface, where the magnetic field is stronger, photons are forced to maintain the polarization mode they had at the emission, due to vacuum birefringence, a QED effect \cite{he36,hs00,hs02,hsl03,fd11,tav+14}, so that the observed polarization degree is little affected by geometrical de-polarization. Hence, magnetar emission is expected to be highly polarized.

With the exception of the first, pioneering observations  by the {\it OSO-8} satellite \cite{weiss+78a,weiss+78b}, no X-ray polarization measurements of neutron stars have been performed till the launch in December 2021 of {\it IXPE} \cite{weiss+22}, a joint NASA-Italian Space Agency (ASI) mission. Each of the three detector units of {\it IXPE} is equipped with a Gas Pixel Detector (GPD) polarimeter, based on the photoelectric effect, a technology which largely increases the sensitivity over older polarimetric facilities. This has allowed the observation of X-ray polarization from several types of astrophysical sources: supernova remnants (SNRs), active galactic nuclei (AGN), pulsars and pulsar wind nebulae (PWNe), black hole binary systems and the Galactic Center, including four magnetar sources. 

In this review we summarize the results and the main physical implications of the {\it IXPE} polarization measurements of the persistent emission from the three AXPs, 4U 0142+61, 1RXS J170849.0--400910 and 1E 2259+586, and from the SGR 1806--20, performed between January 2022 and July 2023. In section \ref{sec:theorypol} we present the basics of polarization in strong magnetic fields. We discuss the spetro-polarimetric models of magnetar emission in section \ref{sec:theorymag}. A short summary of the instrument working principle and capabilities is provided in section \ref{sec:ixpe}, followed by a detailed discussion of the observational results and their theoretical interpretation in section \ref{sec:observations}. We finally draw our conclusions in section \ref{sec:concl}.

\section{Polarization in strong magnetic fields} \label{sec:theorypol}
We summarize in this section, for ease of reading, the basic theory of the polarization of radiation propagating in strong magnetic fields, as well as the radiative processes that determine the photon polarization state at the emission.
\subsection{Polarization properties of the magnetized vacuum and plasma} \label{subsec:vac&plasma}
The polarization state of photons propagating in an ultra-magnetized environment can be derived from the wave equation \cite{lh02,hl03,hl06},
\begin{equation} \label{eqn:waveequation}
\boldsymbol{\nabla}\times(\bar{\boldsymbol{\mu}}\cdot\boldsymbol{\nabla}\times\boldsymbol{E})=\frac{\omega^2}{c^2}\boldsymbol{\epsilon}\cdot\boldsymbol{E}\,,
\end{equation}
where $c$ is the speed of light, $\omega$ and $\boldsymbol{E}$ are the  frequency and electric field vector of the wave, respectively, $\boldsymbol{\epsilon}$ is the dielectric tensor and $\bar{\boldsymbol{\mu}}$ the inverse of the magnetic permeability tensor of the medium in which radiation propagates. Classically, for photons travelling in vacuo, both $\boldsymbol{\epsilon}$ and $\bar{\boldsymbol{\mu}}$ are equal to the unit tensor $\mathbb{I}$. However, strong magnetic fields (like those present around magnetars) can modify the optical properties of the vacuum. Fields in excess of the quantum critical field $B_\mathrm{Q}=4.4\times10^{13}\,\mathrm{G}$ ``polarize'' the virtual electron-positron pairs that populate the vacuum around the source, causing $\boldsymbol{\epsilon}$ and $\bar{\boldsymbol{\mu}}$ to deviate from unity. The components of the dielectric and magnetic permeability tensors in the presence of strong magnetic fields can be written as \cite{kn64,adl71}
\begin{equation}
\begin{array}{l}
\boldsymbol{\epsilon}=\mathbb{I}+\boldsymbol{\epsilon}^{\mathrm{(v)}}=(1+a)\mathbb{I}+q\hat{\boldsymbol{B}}\hat{\boldsymbol{B}} \\
\ \\
\bar{\boldsymbol{\mu}}=\mathbb{I}+\bar{\boldsymbol{\mu}}^{\mathrm{(v)}}=(1+a)\mathbb{I}+m\hat{\boldsymbol{B}}\hat{\boldsymbol{B}}\,,
\end{array}
\end{equation}
with $\hat{\boldsymbol{B}}$ the local magnetic field unit vector. For the typical magnetic field strengths inferred in magnetar candidates ($\approx 10^{14}$--$10^{15}\,\mathrm{G}$), the coefficients $a$, $q$ and $m$ can be approximated as \cite{hl06,fd11}
\begin{equation}
\begin{array}{l}
a=-2\delta \\
q=7\delta \\
m=-4\delta 
\end{array}\,,
\end{equation}
where
\begin{equation}
\delta=\frac{\alpha_\mathrm{F}}{45\pi}\left(\frac{B}{B_\mathrm{Q}}\right)^2\approx 3\times10^{-10}\left(\frac{B}{10^{11}\,\mathrm{G}}\right)^2
\end{equation}
and $\alpha_\mathrm{F}\simeq 1/137$ is the fine-structure constant.

Under the assumption that radiation crosses a region filled by an electron-ion plasma (like a NS atmosphere), in which the main photon-particle interactions are given by collisions and radiative damping, the plasma dielectric tensor can be written as \cite{hl03,hl06}
\begin{equation}
\boldsymbol{\epsilon}^\mathrm{(p)}=\left(\begin{array}{ccc}
\varepsilon & \mathrm{i}g & 0 \\
-\mathrm{i}g & \varepsilon & 0 \\
0 & 0 & \Upsilon
\end{array}\right)\,.
\end{equation}
The parameters $\varepsilon$, $g$ and $\Upsilon$ are defined through the relations \cite{hl06}
\begin{equation}
\begin{array}{ccc}
\varepsilon\pm g &=&1-\dfrac{1}{\omega}\dfrac{\omega_\mathrm{p,e}^2(\omega+\mathrm{i}\nu_\mathrm{ei})+\omega_\mathrm{p,i}^2(\omega+\mathrm{i}\nu_\mathrm{r,e})}{(\omega+\mathrm{i}\nu_\mathrm{r,e}\pm\omega_\mathrm{B,e})(\omega+\mathrm{i}\nu_\mathrm{r,i}\mp\omega_\mathrm{B,i})+\mathrm{i}\omega\nu_\mathrm{ei}} \\
\ &\ & \\
\Upsilon &\simeq& 1-\dfrac{1}{\omega}\left(\dfrac{\omega_\mathrm{p,e}^2}{\omega+\mathrm{i}(\nu_\mathrm{ei}+\nu_\mathrm{r,e})}-\dfrac{\omega_\mathrm{p,i}^2}{\omega+\mathrm{i}(\nu_\mathrm{ei}+\nu_\mathrm{r,i})}\right)\,,
\end{array}
\end{equation}
with $\omega_\mathrm{p,e}$ ($\omega_\mathrm{i,e}$) the electron (ion) plasma frequency, $\omega_\mathrm{B,e}$ ($\omega_\mathrm{B,i}$) the electron (ion) cyclotron frequency, $\nu_\mathrm{r,e}$ ($\nu_\mathrm{r,i})$ the electron (ion) radiation damping frequency and $\nu_\mathrm{ei}$ the electron-ion collision frequency.

In the more general case, photons emitted from the surface of a magnetar will cross a region of space in which both the contributions of the (magnetized) plasma and vacuum are present and is therefore characterized by the tensors
\begin{equation} \label{eqn:totaltensors}
\begin{array}{ll}
\boldsymbol{\epsilon}=\boldsymbol{\epsilon}^\mathrm{(v)}+\boldsymbol{\epsilon}^\mathrm{(p)}&=\left(\begin{array}{ccc}
\varepsilon' & \mathrm{i}g & 0 \\
-\mathrm{i}g & \varepsilon' & 0 \\
0 & 0 & \Upsilon'
\end{array}\right) \\
\ & \\
\bar{\boldsymbol{\mu}}=\mathbb{I}+\bar{\boldsymbol{\mu}}^\mathrm{(v)}\,,
\end{array}
\end{equation}
where, in the low-field limit ($B\ll5\times10^{16}\,\mathrm{G}$), the parameters $\varepsilon'$ and $\Upsilon'$ are given by \cite{hl03}
\begin{equation}
\begin{array}{l}
\varepsilon'=\varepsilon+a \\
\Upsilon'=\Upsilon+a+q\,.
\end{array}
\end{equation}

\subsection{Polarization modes in strong magnetic fields and vacuum resonance} \label{subsec:polmodes}
Considering an orthogonal reference frame $(\boldsymbol{e}_\mathrm{x}\,,\boldsymbol{e}_\mathrm{y}\,,\boldsymbol{e}_\mathrm{z})$, with the $z$ axis along the photon momentum $\boldsymbol{k}$ and the $x$ axis in the plane of  $\boldsymbol{k}$ and the magnetic field direction $\boldsymbol{B}$, the electric field vector can be obtained by solving the wave equation (\ref{eqn:waveequation}) using the expressions (\ref{eqn:totaltensors}) for the dielectric and magnetic permeability tensors and is given by \cite{hl03,hl06}
\begin{equation} \label{eqn:Eunitvector}
\hat{\boldsymbol{E}}=\frac{1}{\sqrt{1+K_\pm^2+K_{\mathrm{z},\pm}}}\left(\begin{array}{c}
\mathrm{i}K_\pm \\
1 \\
\mathrm{i}K_{\mathrm{z},\pm}
\end{array}\right)\,.
\end{equation}
The ellipticity $K=-\mathrm{i}E_\mathrm{x}/E_\mathrm{y}$ for the two general modes $\pm$ in the basis $\boldsymbol{e}_\pm=(\boldsymbol{e}_\mathrm{x}\pm\mathrm{i}\boldsymbol{e}_\mathrm{y})/\sqrt{2}$ can be written as
\begin{equation} \label{eqn:Kpm}
K_\pm=\beta\pm\sqrt{1\pm\beta^2+\frac{m}{1+a}\sin^2\theta_\mathrm{Bk}}\,,
\end{equation}
with
\begin{equation}
\beta=-\frac{\varepsilon'^2-g^2-\varepsilon'\Upsilon'(1+m/a)}{2g\Upsilon'}\frac{\sin^2\theta_\mathrm{Bk}}{\cos\theta_\mathrm{Bk}}\,,
\end{equation}
and $\theta_\mathrm{Bk}$ the angle between $\boldsymbol{k}$ and $\boldsymbol{B}$. Actually, in the reference frame assumed here, the $z$-component in the unit vector $\hat{\boldsymbol{E}}$ translates into an oscillation along the propagation direction $\boldsymbol{k}$ that, considering plane waves, can be neglected \cite{mesz92}.

In normal conditions, when either plasma or vacuum contributions dominate in the dielectric tensor $\boldsymbol{\epsilon}$, it is $|\beta|\gg1$. Using  expression (\ref{eqn:Kpm}) for $K_\pm$ in equation ($\ref{eqn:Eunitvector}$), it follows that photons are polarized in two normal modes, with polarization vectors along
\begin{equation}
\hat{\boldsymbol{E}}_\mathrm{O}=\left(\begin{array}{c}
1 \\ 0 \\ 0
\end{array}\right)\,\,\,\,\,\,\hat{\boldsymbol{E}}_\mathrm{X}=\left(\begin{array}{c}
0 \\ 1 \\ 0
\end{array}\right)\,,
\end{equation}
i.e. in the $\boldsymbol{k}$-$\boldsymbol{B}$ plane and perpendicularly to it, respectively. This two modes are called, respectively, ordinary (O) and extraordinary (X) modes \cite{gp74,mesz92}. In the case of radiation emitted from magnetars, with magnetic field in excess of $B_\mathrm{Q}$, the cross sections of the radiative processes (e.g. scattering and bremsstrahlung) which involve X-mode photons are much suppressed with respect to the ones involving only O-mode ones \cite{her79,vent79,mesz92}. As a consequence, magnetar emission can be reasonably expected to be mostly polarized in the X-mode.

On the other hand, when the plasma and vacuum contributions in the dielectric tensor become comparable  it is $\beta\sim0$, causing the photons to be circularly polarized, so that the distinction between the two normal modes becomes ambiguous \cite{ps79,hl03,lh03}. This occurs at a particular value of the plasma density \cite{lh03,hl03,hl06},
\begin{equation} \label{eqn:vacres}
\rho_\mathrm{V}\simeq 0.964 Y_\mathrm{e}^{-1}\left(\frac{\hbar\omega}{1\,\mathrm{keV}}\right)^2\left(\frac{B}{10^{14}\,\mathrm{G}}\right)^2\lambda^{-2}\,\mathrm{g}\,\mathrm{cm}^{-3}\,,
\end{equation}
with $Y_\mathrm{e}$ the electron fraction of the plasma and $\lambda\simeq1$ a slowly varying function of $B$. Following the evolution of $K_+$ and $K_-$ given in equation (\ref{eqn:Kpm}) as a function of the plasma density, it can be seen that photons should experience complete mode conversion crossing the vacuum resonance density $\rho_\mathrm{V}$, i.e. photons initially polarized in the O-mode may turn into X-mode and vice versa. Nevertheless, assuming photon propagation to be adiabatic, which is verified only for photon energies
\begin{equation}
\mathcal{E}=\hbar\omega\gtrsim\mathcal{E}_\mathrm{ad}=1.49\left(f\left|1-\frac{\omega_\mathrm{B,i}^2}{\omega^2}\right|\tan\theta_\mathrm{Bk}\right)^{2/3}\left(\frac{H_\rho}{5\,\mathrm{cm}}\right)^{-1/3}\,\mathrm{keV}\,,
\end{equation}
where
\begin{equation}
f=\left(\frac{\alpha_\mathrm{F}\beta^2}{15\pi(q+m)}\right)^{1/2}
\end{equation}
and $H_\rho$ is the density scale-height, mode conversion occurs with a probability $P_\mathrm{conv}=1-P_\mathrm{J}$, with \cite{lh03,hl03,hl06}
\begin{equation}
P_\mathrm{J}=\exp\left[-\frac{\pi}{2}\left(\frac{\mathcal{E}}{\mathcal{E}_\mathrm{ad}}\right)^3\right]\,.
\end{equation}
Hence, in the general case mode conversion is only partial and the fraction of extraordinary photons may be even heavily reduced, depending on both the photon energy $\mathcal{E}$ and direction $\boldsymbol{k}$ with respect to the magnetic field $\boldsymbol{B}$. However, for parameter values typical of magnetars, it turns out $P_\mathrm{J}\approx0.03$ for $\mathcal{E}=1.3\mathcal{E}_\mathrm{ad}$, so that, in most of the soft X-ray band (basically above $2$ keV), extraordinary photons can be anyway expected to dominate \cite{kel+24}.

\subsection{Polarization transport in highly magnetized media} \label{subsec:poltransp}
As photons leave the stellar surface, plasma contributions in the dielectric tensor rapidly drop off and, as soon as $\rho\ll\rho_\mathrm{V}$, the polarization properties of radiation 
can be studied solving the wave equation in the limit of zero plasma density  \cite{zhel96,fd11}. Writing the electric field as 
\begin{equation}
\boldsymbol{E}=\boldsymbol{E}_0(z)e^{-\mathrm{i}\omega t}=\boldsymbol{A}(z)e^{\mathrm{i}(k_0z-\omega t)}\,,
\end{equation}
where $k_0=\omega/c$ is the photon wave number and $\boldsymbol{A}=(a_\mathrm{x}e^{\mathrm{i}\varphi_\mathrm{x}},a_\mathrm{y}e^{\mathrm{i}\varphi_\mathrm{y}},a_\mathrm{z}e^{\mathrm{i}\varphi_\mathrm{z}})$ is the electric field complex amplitude, equation (\ref{eqn:waveequation}) can be reduced to a much simpler system of differential equations \cite{fd11,tav+14}
\begin{equation} \label{eqn:diffeqsys}
\begin{array}{ll}
\dfrac{dA_\mathrm{x}}{dz}&=\dfrac{\mathrm{i}k_0\delta}{2}[MA_\mathrm{x}+PA_\mathrm{y}] \\
\ & \\
\dfrac{dA_\mathrm{y}}{dz}&=\dfrac{\mathrm{i}k_0\delta}{2}[PA_\mathrm{x}+NA_\mathrm{y}] \\
\ & \\
A_\mathrm{z}&=-\dfrac{\epsilon_\mathrm{zx}}{\epsilon_\mathrm{zz}}A_\mathrm{x}-\dfrac{\epsilon_\mathrm{zy}}{\epsilon_\mathrm{zz}}A_\mathrm{y}\,.
\end{array}
\end{equation}
The quantities $M$, $N$ and $P$ which appear in equations (\ref{eqn:diffeqsys}) depend on the components of the magnetic permeability tensor, as well as on the magnetic field direction and strength
\begin{equation}
\begin{array}{ll}
M &=\dfrac{(7\hat{B}_\mathrm{x}^2+4\hat{B}_\mathrm{y}^2)\bar{\mu}_\mathrm{xx}-12\delta\hat{B}_\mathrm{x}^2\hat{B}_\mathrm{y}^2}{\hat{\mu}_\mathrm{xx}\hat{\mu}_\mathrm{yy}-16\delta^2\hat{B}_\mathrm{x}^2\hat{B}_\mathrm{y}^2}  \\
\ & \\
N &=\dfrac{(4\hat{B}_\mathrm{x}^2+7\hat{B}_\mathrm{y}^2)\bar{\mu}_\mathrm{yy}-12\delta\hat{B}_\mathrm{x}^2\hat{B}_\mathrm{y}^2}{\hat{\mu}_\mathrm{xx}\hat{\mu}_\mathrm{yy}-16\delta^2\hat{B}_\mathrm{x}^2\hat{B}_\mathrm{y}^2}  \\
\ & \\ 
P &=\dfrac{[3\hat{\mu}_\mathrm{xx}-4\delta(7\hat{B}_\mathrm{y}^2+4\hat{B}_\mathrm{x}^2)]\hat{B}_\mathrm{x}\hat{B}_\mathrm{y}}{\hat{\mu}_\mathrm{xx}\hat{\mu}_\mathrm{yy}-16\delta^2\hat{B}_\mathrm{x}^2\hat{B}_\mathrm{y}^2}\,.
\end{array}
\end{equation}
Since, as also mentioned in \S\ref{subsec:polmodes}, the amplitude $A_\mathrm{z}$ of the oscillation along $\boldsymbol{k}$ is much smaller than $A_\mathrm{x}$ and $A_\mathrm{y}$ \cite{fd11,tav+14}, the last of equations (\ref{eqn:diffeqsys}) can be neglected. Moreover, introducing the Stokes parameters \cite{rl79}
\begin{equation} \label{eqn:stokespar}
\begin{array}{ll}
I&=A_\mathrm{x}A_\mathrm{x}^*+A_\mathrm{y}A_\mathrm{y}^*=a_\mathrm{x}^2+a_\mathrm{y}^2 \\
& \\
Q&=A_\mathrm{x}A_\mathrm{x}^*-A_\mathrm{y}A_\mathrm{y}^*=a_\mathrm{x}^2-a_\mathrm{y}^2 \\
& \\
U&=A_\mathrm{x}A_\mathrm{y}^*+A_\mathrm{y}A_\mathrm{x}^*=2a_\mathrm{x}a_\mathrm{y}\cos\left(\varphi_\mathrm{x}-\varphi_\mathrm{y}\right) \\
& \\
V&=\mathrm{i}\left(A_\mathrm{x}A_\mathrm{y}^*-A_\mathrm{y}A_\mathrm{x}^*\right)=2a_\mathrm{x}a_\mathrm{y}\sin\left(\varphi_\mathrm{x}-\varphi_\mathrm{y}\right)\,,
\end{array}
\end{equation}
where a $*$ denotes the complex conjugate, the system (\ref{eqn:diffeqsys}) can be written in the form \cite{tav+14}
\begin{equation} \label{eqn:diffeqsysstokes}
\begin{array}{ll}
\dfrac{dQ}{dz}&=-\dfrac{k_0\delta}{2}\left(2PV\right) \\
& \\
\dfrac{dU}{dz}&=-\dfrac{k_0\delta}{2}\left(N-M\right)V \\
& \\
\dfrac{dV}{dz}&=-\dfrac{k_0\delta}{2}\left[\left(M-N\right)U-2PQ\right]V\,,
\end{array}
\end{equation}
from which it follows that the polarization state of photons propagating in the magnetized vacuum changes along a typical scale-length
\begin{equation} \label{eqn:ellE}
\ell_\mathrm{E}=\frac{2}{k_0\delta}\simeq130\left(\frac{B}{10^{11}\,\mathrm{G}}\right)^{-2}\left(\frac{\mathcal{E}}{1\,\mathrm{keV}}\right)^{-1}\,\mathrm{cm}\,.
\end{equation}
Hence, for photons with a given energy $\mathcal{E}$, the evolution of the polarization state is only determined by the star's magnetic field strength at any given point. Assuming, for simplicity, that the external field is a dipole (i.e. $B\sim r^{-3}$, with $r$ the radial distance from the star), it changes across a length-scale  \cite{fd11,tav+14}
\begin{equation} \label{eqn:ellB}
\ell_\mathrm{B}=\frac{B}{|\boldsymbol{k}\cdot\boldsymbol{\nabla}B|}=\frac{r}{3}\,.
\end{equation}

\begin{figure}[H]
\begin{center}
\includegraphics[width=11.7cm]{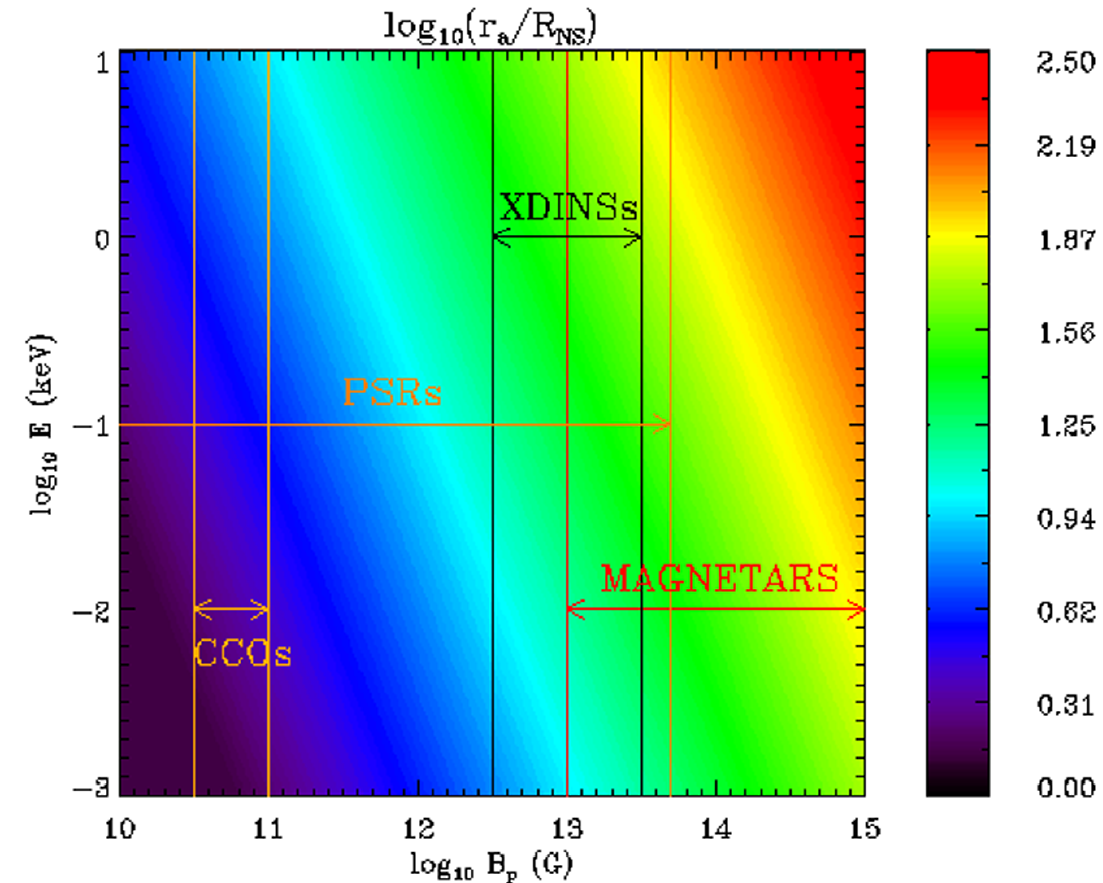}
\caption{Polarization-limiting radius as a function of the photon energy and magnetic field strength. The spin-down magnetic field ranges for different types of NSs are also shown:
radio pulsars (PSRs), central compact objects (CCOs), X-ray dim isolated neutron stars (XDINSs) and magnetars. Figure taken from \cite{tav+15}. \label{fig:rpl}}    
\end{center}
\end{figure}

The way in which the polarization vectors change as radiation propagates can be, therefore, characterized by comparing the two scale-lengths $\ell_\mathrm{E}$ and $\ell_\mathrm{B}$. Close to the stellar surface (i.e. for small values of $r$), $\ell_\mathrm{E}\ll\ell_\mathrm{B}$; as a consequence, the photon polarization vectors change direction much more rapidly than the star's magnetic field. In particular, if a photon has been emitted in the ordinary (extraordinary) mode, it maintains the same mode along its trajectory, as long as $r$ remains sufficiently small. On the other hand, at large distances from the star, it is $\ell_\mathrm{E}\gg\ell_\mathrm{B}$ and the electric field direction remain frozen with respect to that of the magnetic field. This means that, at large distances, the polarization modes can change, potentially washing out the original polarization pattern. 
However, if the transition between these two regions occurs at a large enough distance from the star's surface (so that the magnetic field direction changes more slowly from point to point, see equation \ref{eqn:ellB}), a distant observer may still detect a polarization pattern close to that of radiation at the emission. One can think that the two regions are separated by a sphere of radius $r_\mathrm{pl}$, called the polarization-limiting radius \cite[or also the adiabatic radius,][]{hsl03,tav+15}, where $\ell_\mathrm{E}=\ell_\mathrm{B}$,
\begin{equation} \label{eqn:rpl}
r_\mathrm{pl}=4.8\left(\frac{B_\mathrm{p}}{10^{11}\,\mathrm{G}}\right)^{2/5}\left(\frac{\mathcal{E}}{1\,\mathrm{keV}}\right)^{1/5}\left(\frac{R_\mathrm{NS}}{10\,\mathrm{km}}\right)^{1/5}\,R_\mathrm{NS}\,,
\end{equation}
with $B_\mathrm{p}$ the polar magnetic field strength and $R_\mathrm{NS}$ the NS radius. Although the transition between the two regimes is actually gradual (according to equations \ref{eqn:diffeqsys} and \ref{eqn:diffeqsysstokes}), one can anyway conclude that the larger the polarization limiting-radius the closer the observed polarization properties are to those at the emission.
Given the dependence of $r_\mathrm{pl}$ on the polar magnetic field strength, this especially holds  for highly magnetized neutron stars, such as magnetars (see  Figure \ref{fig:rpl}).

\section{Polarization properties of magnetar X-ray emission} \label{sec:theorymag}

Both the spectral and polarization properties of radiation emitted from magnetars depend on the physical state of the star surface (whether the solid crust is exposed or covered by a thin atmospheric layer), as well as on the interactions photons undergo with charges in the star's magnetosphere. In this section we briefly discuss the main models for thermal emission from magnetars and the role of  magnetospheric effects. 

\subsection{Magnetized atmosphere} \label{subsec:atmo}
The surface of a NS may be covered by a geometrically-thin (typical scale-length $\approx0.1$--$10\,\mathrm{cm}$), optically-thick atmospheric layer. NS atmospheres have been investigated by many authors under different assumptions, concerning the chemical composition \cite[e.g. partially- or fully-ionized hydrogen, helium or other light elements, as well as heavy elements,][]{rom87,shib+92,pavl+94,sul+09} and magnetic effects \cite[including mode conversion at the vacuum resonance][]{lh02,hl03,vl06,hl06}. 

Current modelling of highly magnetized atmospheres assumes both scattering and thermal bremsstrahlung as the main sources of opacity \cite{lloyd03,pot+04,sul+09,pot+14}. As mentioned in \S\ref{subsec:polmodes}, the opacities of X-mode photons are suppressed (by a factor $(B/B_\mathrm{Q})^{-2}$) with respect to the unmagnetized case, causing the emerging radiation to be polarized predominantly in the extraordinary mode, with high polarization degrees \cite{gonz+16,tav+20}. However, this scenario may change due to complete/partial mode conversion at the vacuum resonance, switching the dominant polarization mode from X to O or, in any case, lowering the polarization degree of the emitted radiation. Recent calculations, performed assuming a pure-hydrogen, fully ionized atmosphere, show that the expected polarization pattern strongly depends on the strength of the star's magnetic field. A substantial reduction of the polarization degree at the emission can be expected for moderate magnetic field strengths ($\lesssim5\times 10^{13}\,\mathrm{G}$) at photon energies below $\approx4$--$5\,\mathrm{keV}$; on the other hand, for magnetar-like magnetic fields ($\gtrsim10^{14}\,\mathrm{G}$) the situation is substantially unchanged with respect to that in which vacuum effects are neglected, at least for energies above $2\,\mathrm{keV}$ \cite{kel+24}.

\subsection{Condensed surface} \label{subsec:solid}
Sufficiently strong magnetic fields alter the structure of matter. For $B\gtrsim2.4\times10^9\,\mathrm{G}$ the electron gyroradius becomes smaller than the Bohr radius, and atoms are elongated along the direction of the magnetic field. Then, if the temperature is low enough, molecular chains can form via covalent bonding. As a result, the 
gaseous layer above the surface may experience a phase transition, precipitating onto the crust and leaving exposed the star's condensed surface, a phenomenon known as  magnetic condensation \cite{brink80,tzd04,paz+05,vanad+05,pot+12}. 

\begin{figure}[H]
\begin{center}
\includegraphics[width=14.0cm]{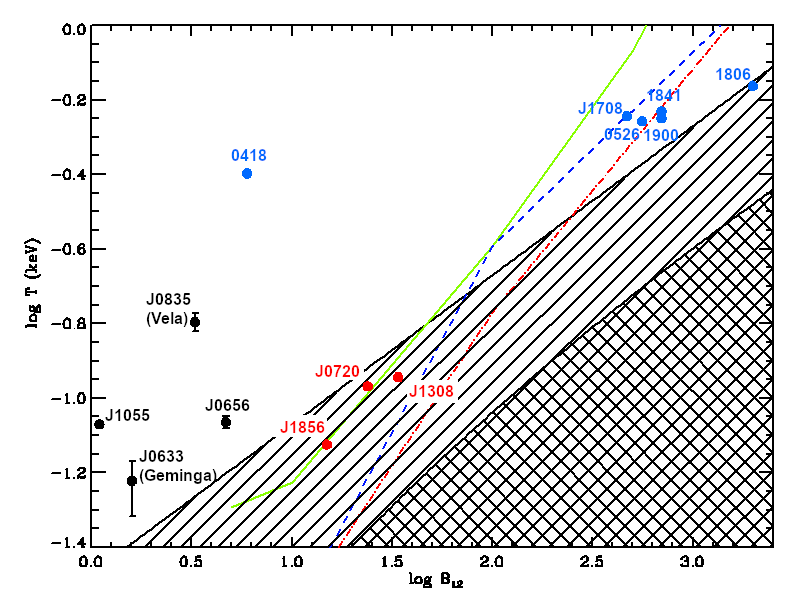}
\caption{Critical temperature $T_\mathrm{crit}$ for magnetic condensation plotted as a function of the surface magnetic field strength $B$ (in units of $10^{12}\,\mathrm{G}$). The values for which magnetic condensation of hydrogen and iron can occur, according to the model by \cite{lai01}, are highlighted by the shaded and hatched zones, respectively. The green-solid, blue-dashed and red-dash dotted lines mark the limit of the regions in which magnetic condensation of iron, carbon and helium, respectively, can occur according to the model by \cite{ml06b,ml07}. The effective temperature and surface magnetic field of some pulsars (PSR J1055--6028, J0633$+$1746, J0656$+$14 and J0835--4510, black filled circles with error bars), XDINSs (RX J1856.5--3754, RX J0720.4--3125 and J1308.6$+$2127, red filled circles) and magnetars (SGR 0418$+$5729, 0526--66, 1806--20 and 1900$+$14; AXP 1RXS J170849.0--400910 and 1E 1841--045, blue filled circles) are also shown \cite{tur09,ok14}. Figure adapted from \cite{tav+20}. \label{fig:tcrit}}    
\end{center}
\end{figure}

The critical temperature $T_\mathrm{crit}$ at which this transition occurs depends in general on both the magnetic field strength and the chemical composition, as shown in Figure \ref{fig:tcrit}. By comparing the values of $T_\mathrm{crit}$ obtained with different  models \cite{lai01,ml06a,ml06b,ml07} with the effective temperature derived from observations of several isolated NSs, information about the physical state of the outermost stellar layers can be obtained. Although the theoretical determination of the critical temperature is still affected by large uncertainties, the extremely strong fields and moderate temperatures make magnetars the best candidates for magnetic condensation. The spectrum emitted by the condensate is thermal and blackbody-like above the electron plasma frequency, not much different from what one expects to observe in the case of radiation emitted by a magnetized atmosphere \cite{pot+12,tav+20}. In this respect, X-ray polarimetry may provide an independent way to disentangle the two emission models. In the $2$--$10\,\mathrm{keV}$ band (which is the one accessible to current instrumentation), radiation emitted from the bare, condensed surface is expected to be only mildly polarized ($\lesssim30\%$), with either O- or X-mode dominating depending on both the photon energy and propagation direction with respect to the star magnetic field. This is at variance with what is predicted for radiation coming from an atmospheric layer (see \S\ref{subsec:atmo}), with quite large polarization ($\gtrsim 70$--$80\%$) and dominated by the X-mode (even accounting for vacuum resonance effects).

In the following we will discuss the case of condensed surface radiation in two  limiting cases \cite{vanad+05,pot+12}:
\begin{itemize}
\item fixed-ions, in which only the electrons may freely respond to incoming electromagnetic waves, while ions are considered to be fixed in the lattice;
\item free-ions, in which both electron and ion motions in response to an incoming electromagnetic wave are considered.
\end{itemize}
Although the real situation should lie in between these two limits, a general and self-consistent description is still lacking; so we will refer to these two cases separately.

\subsection{Resonant Compton scattering in magnetar magnetospheres} \label{subsec:RCS}

According to the twisted-magnetosphere model \cite{tlk02}, some of the helicity of the internal magnetic field of magnetars can be transferred to the external one following the displacements of single surface elements onto which the external field lines are anchored. This makes the external field non-potential, requiring that charged particles flow along the closed field lines to sustain the twist. Although it is more likely that the twist involves a small bundle of external field lines \cite{belo09}, with both electrons and positrons contributing to the magnetospheric currents \cite{bt07}, here we resort to a simplified model with a globally-twisted dipole field and currents made  by electrons and ions lifted from the stellar surface, which nevertheless proved capable to explain the available spectral observations \cite{ntz08,kraw+22}. Under these assumptions, the polar components of the external field, $(B_r,B_\theta,B_\phi)$, with $\theta$ and $\phi$ the magnetic colatitude and azimuth, respectively, can be written as \cite{tlk02,pav+09,fd11}
\begin{equation} \label{eqn:twistedB}
\boldsymbol{B}=\frac{B_\mathrm{p}}{2}\left(\frac{r}{R_\mathrm{NS}}\right)^{-2-p}\boldsymbol{\varrho}=\frac{B_\mathrm{p}}{2}\left(\frac{r}{R_\mathrm{NS}}\right)^{-2-p}\left[
-\dfrac{df}{d\cos\theta},\dfrac{pf}{\sin\theta},\sqrt{\dfrac{pC(p)}{p+1}}\dfrac{f^{1+1/p}}{\sin\theta}
\right]\,,
\end{equation}
where $f$ is a function of $\cos\theta$, satisfying the Grad-Shafranov equation and $C(p)$ is an eigenvalue which depends only on the radial index $p$. Actually, a convenient way to characterize the twist is by evaluating the angular displacement between the field line footpoints in the northern and southern magnetic hemispheres, i.e. the twist angle
\begin{equation} \label{eqn:deltaphi}
\Delta\phi_\mathrm{N-S}=\lim_{\theta\rightarrow0}\int_\theta^{\pi/2}\frac{B_\phi}{B_\theta\sin\theta}d\theta\,;
\end{equation}
$\Delta\phi_\mathrm{N-S}$ is a function of $p$ only, so the amount of twist is directly related to the radial index.
By imposing that the current density along the closed field lines satisfies the Maxwell equation $\boldsymbol{j}=c\boldsymbol{\nabla}\times\boldsymbol{B}/4\pi$, the electron density in the magnetosphere can be expressed as \cite{tlk02,ntz08}
\begin{equation} \label{eqn:ne}
n_\mathrm{e}=\frac{p+1}{4\pi e}\left(\frac{B_\phi}{B_\theta}\right)\frac{B}{r|\langle\bar{\beta}\rangle|}\,,
\end{equation}
with $e$ and $\langle\bar{\beta}\rangle$ the electron charge and velocity (in unit of the speed of light $c$) along the closed field lines, respectively.

At the cyclotron resonance ($\omega=\omega_\mathrm{B}$ in the particle rest frame), the electron scattering cross sections become sufficiently large to make the magnetosphere optically thick for resonant Compton scattering (RCS). Upscattered photons populate a  power-law tail above $\approx 3$--$4\,\mathrm{keV}$, in agreement with what is observed in persistent magnetar spectra 
\cite{ntz08,fd11,tav+14}. Resonant cross-sections are mode-sensitive
\begin{equation} \label{eqn:RCScrosssection}
\begin{array}{ll}
\sigma_\mathrm{OO}&=\dfrac{1}{3}\sigma_\mathrm{OX} \\
\ & \\
\sigma_\mathrm{XX}&=3\sigma_\mathrm{XO}\,,
\end{array}
\end{equation}
where the first (second) subscript refers to the incoming (outgoing) photon. As a result, for saturated RCS, one can expect that radiation emerging from the magnetosphere is polarized predominantly in the X-mode, with a polarization fraction $\approx 33\%$ \cite{tav+20}.

For values  typical of magnetars, the ratio between the plasma and vacuum contributions in the dielectric tensor is (\ref{eqn:totaltensors}) turns out to be \cite{zhel96,fd11}
\begin{equation}
\frac{(\omega_\mathrm{p,e})^2}{\delta}\approx10^{-7}\sin^2\theta\Delta\phi_\mathrm{N-S}\left(\frac{B_\mathrm{p}}{10^{14}\,\mathrm{G}}\right)^{-1}\left(\frac{R_\mathrm{NS}}{10^6\,\mathrm{cm}}\right)^{-1}\left(\frac{r}{R_\mathrm{NS}}\right)^{1+p}\left(\frac{\mathcal{E}}{1\,\mathrm{keV}}\right)^{-2}\,.
\end{equation}
Moreover, RCS can occur within a distance \cite{fd11}
\begin{equation}
r_\mathrm{esc}\approx10\left[\frac{\varrho}{2}(1-\cos^2\theta_\mathrm{Bk})^{-1/2}\left(\frac{B_\mathrm{p}}{10^{14}\,\mathrm{G}}\right)\left(\frac{\mathcal{E}}{1\,\mathrm{keV}}\right)^{-1}\right]^{1/(2+p)}R_\mathrm{NS}\,,
\end{equation}
which is much smaller than the polarization-limiting radius $r_\mathrm{pl}$ (see equation \ref{eqn:rpl} and Figure \ref{fig:rpl}). Therefore, the polarization mode evolution across the magnetosphere can be anyway considered as solely determined by the magnetized vacuum, taking for the polarization pattern at emission that of radiation emerging from $r_\mathrm{esc}$.

\section{X-ray polarimetry with {\em IXPE}} \label{sec:ixpe}

\begin{figure}[h]
\begin{center}
\includegraphics[width=9.7cm]{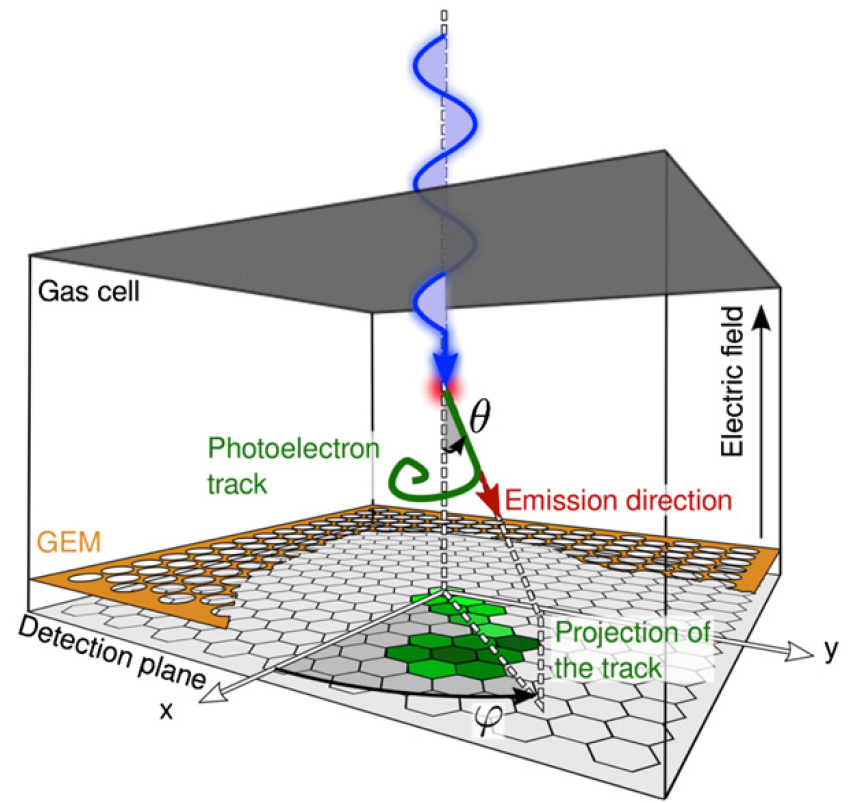}
\caption{The illustration shows the main components of the GPD on board of {\it IXPE} (see text for details). Figure taken from \cite{mul14}. \label{fig:gpd}}    
\end{center}
\end{figure}

The Imaging X-ray Polarimetry Explorer \cite[\textit{IXPE}][]{weiss+22} is a NASA SMEX mission, developed in cooperation with the Italian Space Agency (ASI), which carried into orbit the first X-ray polarimeter more than 40 years after the last observatory capable to perform polarization measurements in the soft X-rays, the \textit{OSO-8} satellite \cite{weiss+76}. The previous instrumentation was based on Bragg diffraction and Compton scattering; {\it IXPE} instead exploits the photoelectric effect \cite{mul14}. The core of the instrument is the Gas Pixel Detector \cite[GPD,][]{cos+01,bell+03}. The working principle is sketched in Figure \ref{fig:gpd}: the detector is composed of a cell filled by a mixture of $20\%$ helium and $80\%$ dimethyl ether. The low extraction potential of the gas atoms facilitates the extractions of photo-electrons by the incoming X-rays, which enter the cell through a thin beryllium  window. Under an applied voltage, each photo-electron drifts to a Gas Electron Multiplier (GEM), before being collected at the pixel anode.

Assuming that the incoming photon is absorbed by a spherically symmetric shell of the gas particles, the (differential) photoelectric cross section turns out to be \cite{heit54,mul14}
\begin{equation} \label{eqn:photoelcrosssec}
\frac{d\sigma}{d\Omega}=r_0^2\alpha_\mathrm{F}Z^5\left(\frac{m_\mathrm{e}c^2}{\mathcal{E}}\right)^{7/2}\frac{4\sqrt{2}\sin^2\theta}{(1-\bar{\beta}\cos\theta)^4}\cos^2\phi\,;
\end{equation}
here $\theta$ is the angle between the photo-electron momentum and the incoming photon direction and $\phi$ is the associated azimuth, 
counted from the photon electric field direction, $d\Omega=d(\cos\theta)d\phi$, $r_0$ is the classic electron radius, $m_\mathrm{e}$ the electron mass, $Z$ the atomic number of the absorber and $\bar{\beta}$ the photo-electron velocity, in units of the speed of light. From equation (\ref{eqn:photoelcrosssec}) it is clear that the photo-electron is most likely emitted along the polarization direction of the incoming photon ($\phi=0$). Starting from the photoelectric cross section, the number of photo-electrons emitted at a given azimuthal angle $\phi$ can be written as \cite{mul14}
\begin{equation} \label{eqn:modulation}
\mathcal{N}(\phi)=\mathcal{N}_\mathrm{tot}\left[\mathcal{P}\frac{\cos^2\phi}{\pi}+\frac{1-\mathcal{P}}{2\pi}\right]\,,
\end{equation}
where $\mathcal{N}_\mathrm{tot}$ is the total number of events and $\mathcal{P}$ is the ratio of events due to polarized photons $\mathcal{P}=\mathcal{N}_\mathrm{pol}/\mathcal{N}_\mathrm{tot}$, which is defined as the polarization degree (PD) of the incoming radiation. Equation (\ref{eqn:modulation}) shows that the expected response of the detector to polarized radiation is a cosine squared modulation, 
\begin{equation} \label{eqn:modulation_general}
\mathcal{M}(\varphi)=K+A\cos^2(\varphi-\varphi_0)\,,
\end{equation}
with $\varphi$ the azimuth of the photoelectron track counted from the reference direction in the detector (see Figure \ref{fig:gpd}). 
Hence, the polarization degree is related to the modulation amplitude $A$, while the angle $\varphi_0$, which corresponds to the azimuthal displacement between the incoming photon polarization direction and the detector reference axis, identifies the polarization angle (PA) of the collected radiation.

For incident $100\%$ polarized radiation (i.e. $\mathcal{P}=1$) the amplitude $A$ of the signal defines the modulation factor $\mu$ of the instrument. So, the polarization degree can be obtained from the modulation amplitude normalized to $\mu$,
\begin{equation} \label{eqn:PDnorm}
\mathrm{PD}=\frac{M}{\mu}\,\,\,\,\,\mathrm{where}\,\,\,\,\,M=\frac{\mathcal{M}_\mathrm{max}-\mathcal{M}_\mathrm{min}}{\mathcal{M}_\mathrm{max}+\mathcal{M}_\mathrm{min}}\,.
\end{equation}
On the other hand, for completely unpolarized radiation ($\mathcal{P}=0$), the detector response as a function of the azimuthal angle is expected to be flat. However, essentially for statistical reasons, the instrumental response function can be always fitted by a cosine squared modulation, which is positive definite, yielding a non-zero polarization degree even in the cases of unpolarized photons. This instrumental limitation is quantified by the Minimum Detectable Polarization (MDP), which corresponds to the amplitude of the modulation curve above which the incoming radiation is polarized at a given level of confidence. In analyzing \textit{IXPE} data, the MDP at $99\%$ confidence level is used, which is defined as \cite{weiss+10,sk13}
\begin{equation} \label{eqn:MDP99}
\mathrm{MDP}_{99}=\frac{4.29}{\eta\mu\mathcal{S}}\sqrt{\frac{\mathcal{B}+\eta\mathcal{S}}{\mathcal{A}t_\mathrm{exp}}}\,,
\end{equation}
where $\eta$ is the instrumental efficiency, $\mathcal{A}$ is the collecting area of the detector, $\mathcal{S}$ ($\mathcal{B}$) is the signal (background) count rate and $t_\mathrm{exp}$ the exposure time. A measure providing $\mathrm{PD_\mathrm{obs}}>\mathrm{MDP}_{99}$ means that radiation with polarization degree $\mathrm{PD_\mathrm{obs}}$ has been detected at $99\%$ confidence level.

The polarization degree and angle can be also derived from the Stokes parameters (equations \ref{eqn:stokespar}), as \cite{rl79}
\begin{equation} \label{eqn:PDPA}
\mathrm{PD}=\frac{\sqrt{Q^2+U^2}}{I}\,\,\,\,\,\,\,\,\,\,\,\mathrm{PA}=\frac{1}{2}\arctan\left(\frac{U}{Q}\right)\,.
\end{equation}
Hence, a polarization measurement with the GPD provides an estimate of the Stokes parameters as well. By comparing equations (\ref{eqn:modulation_general}), (\ref{eqn:PDnorm}) and (\ref{eqn:PDPA}), the latter are expressed as functions of the modulation curve parameters as
\begin{equation} \label{eqn:measstoekspar}
\begin{array}{rl}
Q&=\dfrac{A\cos(2\varphi_0)}{2\mu}  \\
\ & \\
U&=\dfrac{A\sin(2\varphi_0)}{2\mu}\,. 
\end{array}
\end{equation}

\begin{figure}[h]
\begin{center}
\includegraphics[width=12.7cm]{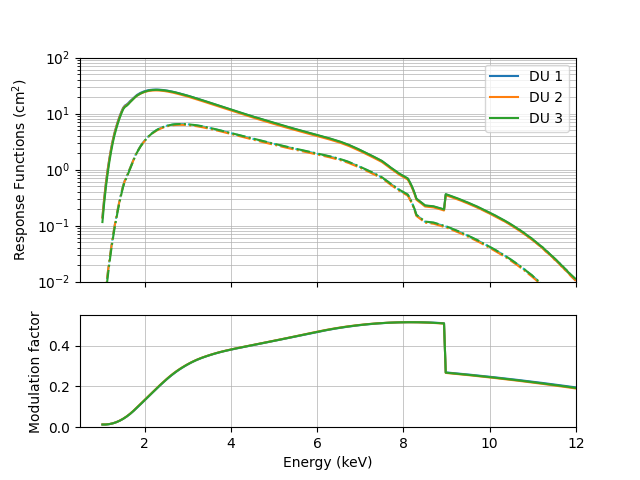}
\caption{Top panel: the effective area (solid lines) and modulation response function (dash-dotted lines) of {\it IXPE} DU 1 (cyan), 2 (orange) and 3 (green) for a point-like, on-axis source at infinity, plotted as a function of the photon energy (notice that the curves are almost superimposed). Bottom panel: the correspondent modulation factors $\mu$ (note that all curves are superimposed). \label{fig:effarea}}    
\end{center}
\end{figure}

The {\textit IXPE} observatory comprises three identical X-ray telescopes, each one equipped with its own GPD and mirror module assembly. They form together the three detector units (DUs) of the instrument, which operate in the $2$--$8\,\mathrm{keV}$ energy band. The effective area is basically the same for all the three DUs (with that of DU 2 only $1$--$2\%$ lower than the DU 1 and DU 3 ones). In particular, the effective area peaks at $2.26\,\mathrm{keV}$, attaining a value $\approx 26\,\mathrm{cm}^2$, then decreasing monotonically in the {\it IXPE} working energy band. In particular, it is $\approx 11\,\mathrm{cm}^2$ at $4\,\mathrm{keV}$ and $\approx 1.4\,\mathrm{cm^2}$ at $7.5\,\mathrm{keV}$. Figure \ref{fig:effarea} shows the energy-dependent behavior of the effective area and modulation response function for each of the 3 {\it IXPE} DUs, together with the correspondent modulation factor $\mu$ (given by the ratio between modulation response function and effective area). The latter turns out to be $\approx 0.2$ at $2.26\,\mathrm{keV}$ and increases monotonically between $2$ and $8\,\mathrm{keV}$, attaining values $\approx0.4$ at $4\,\mathrm{keV}$ and $\approx0.5$ at $7.5\,\mathrm{keV}$. The instrumental response functions are publicly available, and provided in the {\it IXPE} calibration Database\footnote{See the on line CALDB at \url{https://heasarc.gsfc.nasa.gov/docs/ixpe/caldb}}.

\section{{\it IXPE} magnetar observations} \label{sec:observations}

Since its launch, on December 9, 2021, {\it IXPE} observed about 60 targets belonging to different classes (Galactic black hole and NS sources, supernova remnants, AGNs, etc.)
offering an unprecedented view of the X-ray universe. In this section we discuss the X-ray polarization measurements of the four magnetar sources performed during the first two years of operations.

\subsection{AXP 4U 0142+61} \label{subsec:4u}
The AXP 4U 0142+61 is the brightest among the magnetar candidates, located in  Cassiopeia  (R.A. $01^\mathrm{h}$ $46^\mathrm{m}$ $22^\mathrm{s}.41$, DEC. $61^\circ\,45'\,03''.2$) at a distance of $3.6\,\mathrm{kpc}$ \cite{rea+07,tav+22}. It has been observed by {\it IXPE} from January 31 to February 27, 2022, for a total exposure time of $t_\mathrm{exp}=840\,\mathrm{ks}$. Background and spurious modulation contributions have been subtracted from level 2 data according to the prescribed procedures \cite{bald+22,dimar+23}. 

The timing analysis returned a value of spin frequency $f=0.115079336\pm6\times 10^{-9}\,\mathrm{Hz}$ and the long observation allowed to derive a significant estimate of the frequency derivative,  $\dot{f}=-(2.1\pm0.7)\times10^{-14}\,\mathrm{Hz\,s}^{-1}$ (MJD epoch 59624.050547); these values are in agreement (within the errors\footnote{Here and in the following errors are given at $1\sigma$ confidence level, unless explicitly stated otherwise.}) with previous estimates of spin period and spin-down rate \cite{ok14,dk14}, implying a dipolar magnetic field strength at the surface $B\approx1.5\times10^{14}\,\mathrm{G}$. 

\begin{figure}[h]
\begin{center}
\includegraphics[width=13.7cm]{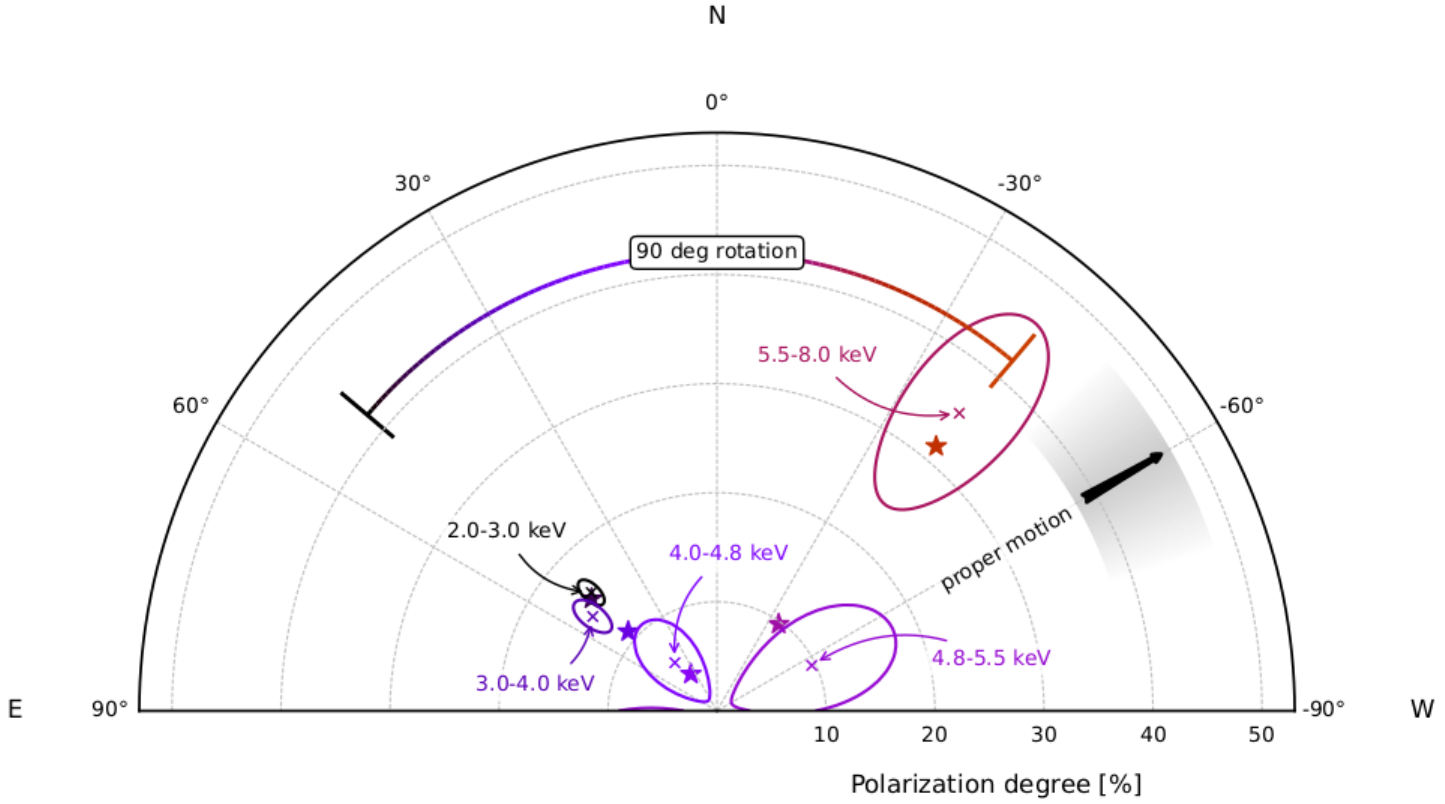}
\caption{Polar plot showing the phase-integrated, energy-dependent PD and PA (crosses with $1\sigma$ confidence contours) measured by {\it IXPE} for the AXP 4U 0142+61. The results of a numerical simulation obtained assuming that the magnetar thermal emission from an equatorial belt on the star condensed surface is reprocessed by RCS in the magnetosphere are marked by stars. Dashed lines indicate the values of PA at low ($2$--$4\,\mathrm{keV}$) and high ($5$--$8\,\mathrm{keV}$) energies. The proper motion direction of the pulsar in the plane of the sky, as measured by \cite{tend+15}, is also shown (black arrow) with its uncertainty (gray-shaded region). Figure taken from \cite{tav+22}. \label{fig:endeppol4U}}    
\end{center}
\end{figure}

A phenomenological fit of the {\it IXPE} spectrum in the $2$--$8\,\mathrm{keV}$ energy range using XSPEC \cite{arn96} revealed the presence of two components, either two blackbodies (BBs) or a BB and power-law (PL). {\it IXPE} lack of sensitivity below $2\,\mathrm{keV}$ makes it impossible to constrain the interstellar absorption, so the hydrogen column density $N_\mathrm{H}$ was frozen to the most recent value found in literature \cite{denh+08}. Both models yield acceptable fits with comparable statistical 
significance. The BB+PL decomposition returned parameter values compatible with those obtained from {\it XMM-Newton} data  \cite{rea+07,denh+08}, with BB temperature $kT_\mathrm{BB}=0.471\pm0.004\,\mathrm{keV}$ and photon index $\Gamma=3.69\pm0.05$ \cite{tav+22}. 

\begin{figure}[h]
\begin{center}
\includegraphics[width=13.7cm]{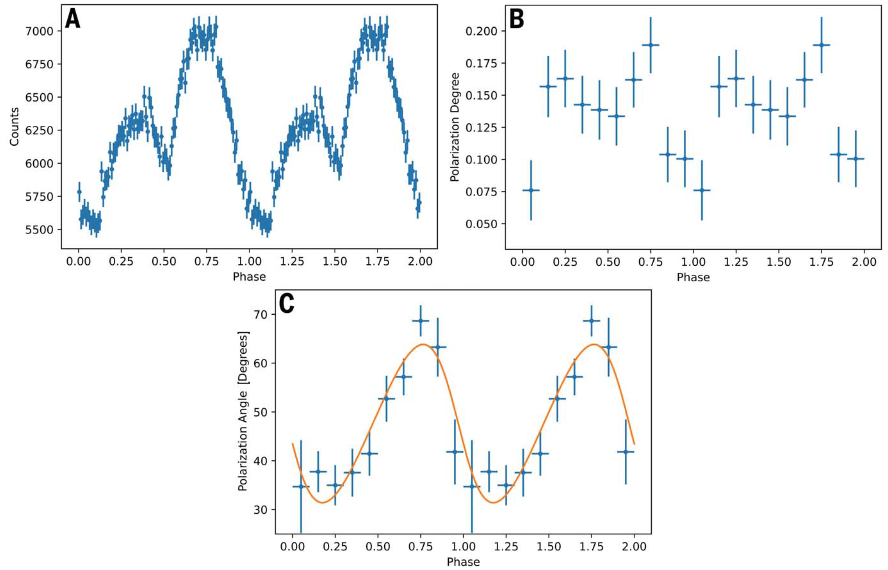}
\caption{Energy-integrated ($2$--$8\,\mathrm{keV}$), phase-dependent measurement of the persistent radiation from the AXP 4U 0142+61 by {\it IXPE}. Cyan filled circles represent the flux (panel A), PD (panel B) and PA (panel C); error bars are at $1\sigma$ confidence level for the flux, while they denote the $\Delta\log L=1$ uncertainty for PD and PA, where $L$ is the unbinned likelihood \cite{gonz+23}. The orange solid line in panel C represents the best fitting rotating vector model (see text for more details). Figure taken from \cite{tav+22}. \label{fig:phdeppol4U}}    
\end{center}
\end{figure}

Joining together the phase- and energy-integrated contributions of the three DUs in the entire $2$--$8\,\mathrm{keV}$ {\it IXPE} band, the Stokes parameters turn out out be  $Q/I=0.013\pm0.008$ and $U/I=0.120\pm0.008$. This translates into $\mathrm{PD}=13.5\pm0.8\%$,  well above the $\mathrm{MDP}_\mathrm{99}\approx2\%$, and $\mathrm{PA}=48.5^\circ\pm1.6^\circ$ measured East of the celestial North; the significance of the detection is  $\approx 17\sigma$. The phase-averaged polarization resolved into five energy bins shows that 
the polarization degree decreases from $15.0\pm1.0\%$ ($\mathrm{MDP}_{99}\approx4\%$) at $2$--$3\,\mathrm{keV}$ down to be compatible with zero at $4$--$5\,\mathrm{keV}$. Then it increases again, attaining a value $35.2\pm7.1\%$ ($\mathrm{MDP}_{99}\approx21\%$) at $6$--$8\,\mathrm{keV}$. At the same time, the measured polarization angle swings from $\approx50^\circ$ at $2$--$4\,\mathrm{keV}$ to $\approx-40^\circ$ at $5$--$8\,\mathrm{keV}$ (see Figure \ref{fig:endeppol4U}). 


The polarization degree attained at high energies ($\approx 35\%$) is compatible, within the errors, with the value expected by RCS (see \S\ref{subsec:RCS}, equations \ref{eqn:RCScrosssection}). An interpretation of the X-ray spectrum in terms of a BB+PL model is also in agreement with the RCS paradigm and this supports a picture in which the $6$-$8\,\mathrm{keV}$ radiation is dominated by X-mode photons. On the other hand, the swing in the polarization angle, by exactly $90^\circ$ between the low- and high-energy ranges, argues in favor of an excess of O-mode photons at lower energies. 

Emission coming from a magnetized, light-element atmosphere hardly produces such a pattern, if the magnetic field is that derived from spin-down ($B\approx 1.5\times10^{14}\,\mathrm{G}$) even considering vacuum effects in the plasma \cite{kel+24}. On the other hand, dominant O-mode photons could be explained if the condensed surface of the star is exposed \cite[and the emitting region is observed under a favourable geometry, see][]{tav+20}. Actually, in order to produce a polarization degree as high as that observed by {\it IXPE} between $2$ and $4\,\mathrm{keV}$ ($\lesssim15\%$), radiation should not come from the entire NS surface, but rather from a zone limited to the (magnetic) equatorial region. Reprocessing of radiation in the star's magnetosphere is able to produce the $90^\circ$ polarization angle swing at the observed energies, allowing the polarization degree to increase up to $\approx33\%$ at $\sim8\,\mathrm{keV}$. Results of a simulation based on this picture using the Monte Carlo code discussed in \cite{tav+14,tav+20}
are shown in Figure \ref{fig:endeppol4U}. 
The computed polarization observables fall basically inside the $1\sigma$ confidence contour of the {\it IXPE} measurements, confirming the plausibility of the model.
An atmosphere layer with an inverted temperature gradient (i.e. where the temperature increases at lower optical depths) can also produce radiation which is O-mode dominated and 
becomes X-mode dominated upon RCS in the magnetosphere. 
An inverted temperature gradient can be produced by heating from the returning particles which flow along the closed field lines. However, no numerical model exists as yet to allow a direct comparison with the observed data in this case.

Alternatively, high-energy X-rays can be taken to be O-mode polarized, with an excess of X-mode photons at lower energies. This pattern can be in principle achieved if thermal radiation from the cooling surface is firstly reprocessed by unsaturated thermal Compton scattering in a plasma layer close to the star crust and then by saturated comptonization in a sort of magnetar corona. Another possible scenario was put forward by \citet{lai23}, who explained the polarization angle swing at $4$--$5\,\mathrm{keV}$ in terms of partial mode conversion at the vacuum resonance in a magnetized atmosphere. This interpretation was also motivated by the fact that the pulsar proper motion in the plane of the sky turns out to be oriented at $60^\circ\pm12^\circ$ West of North \cite{tend+15}, which is $\approx 20^\circ$ apart from the polarization direction of photons at high energies. Since radio polarization measurements in some pulsars \cite{janka+22} show an alignment between the proper motion and the spin axis projected on the observer's sky, one may argue that high-energy photons are predominantly polarized in the O-mode. This is because PA should be $0^\circ$ ($90^\circ$) for radiation mainly polarized in the O-mode (X-mode) taking as a reference the spin axis projection onto the plane of the sky \cite{tav+15}. However, we notice that the association of the spin axis projection with the proper motion is still debated and, in some cases, a spin axis perpendicular to the proper motion must be assumed to explain the NS formation mechanism in a way coherent with observations \cite{cw02}. The model discussed in \cite{lai23} is still preliminary and can not explain at the same time the PA swing and the mild PD observed at low and high energies, nor the presence of a PL tail. Moreover, recent detailed calculations do not support an atmosphere model for 4U 0142+61  \cite[][]{kel+24}

The (energy integrated) flux pulse profile observed by {\it IXPE} is double-peaked, in agreement with that previously reported by Rea et al. \cite{rea+07}. Its shape may result from  both the limited extension of the emission region (which is an equatorial belt in the model discussed earlier) and by the anisotropy of the emission itself \cite[as it is the case of the condensed surface, see][]{pot+12,tav+20}. The behaviour of polarization degree in phase resembles that of the flux, with the phase  of the two peaks basically coincident with those of the light curve (see Figure \ref{fig:phdeppol4U}A and B). This is indicative that the observed polarization pattern bears the imprint of that at the emission (i.e. on the surface and across the magnetosphere). 
On the other hand, the sinusoidal behavior of the polarization angle (Figure \ref{fig:phdeppol4U}C) appears to be completely uncorrelated with both the flux and the polarization degree. Such a behavior is expected in pulsars, where pulsations originate from very small regions close to the star's surface. Under these conditions, the phase-dependent polarization angle is expected to follow the rotating vector model \cite[RVM,][]{rc69}; for slowly rotating NSs (i.e. in the non-relativistic limit), the expression that relates PA to the rotational phase $\gamma$ can be written as \cite{pout20}
\begin{equation} \label{eqn:RVM}
\tan(\mathrm{PA})=\frac{\sin\xi\sin\gamma}{\cos\chi\sin\xi\cos\gamma-\sin\chi\cos\xi}\,,
\end{equation}
where the angle $\chi$ and $\xi$ are the inclinations of the LOS and the magnetic dipole axis with respect to star's spin axis, respectively. However, in the case of magnetars this relation does not necessarily hold, due to both the larger extension of the emitting region on the surface (even assuming radiation coming from an equatorial belt, as discussed above) and the departures from a purely dipolar topology of the surface magnetic field. It can be anyway shown \cite{tav+22} that the polarization angle behavior as a function of the rotational phase again follows equation (\ref{eqn:RVM}) at a large enough distance from the star, where the magnetic field can be assumed reasonably as dipolar. This apparent contradiction (PD fixed by the properties at the surface and PA determined at large distances) can be explained by vacuum birefringence. In fact, as pointed out in \S\ref{subsec:poltransp}, in a strongly magnetized vacuum the photon polarization vectors continuously change in direction close to the star's surface to keep pace with that of the magnetic field, freezing only at a distance $\approx r_\mathrm{pl}$ (which for magnetars is rather large). On the other hand, it is this same mechanism that ensures that the polarization properties at emission are preserved up to great distances from the source. Hence, it can be argued that the peculiar behavior observed with {\it IXPE} provides at least a first hint that vacuum birefringence effects are at work around 4U 0142+61.

\subsection{AXP 1RXS J170849.0--4009100} \label{subsec:1708}
The AXP 1RXS J170849.0--4009100 (1RXS J1708 hereafter, for short) is the most strongly polarized source discovered so far by {\it IXPE} \cite{zane+23}. Observed between September 19 and October 8, 2022, for a total on source time $t_\mathrm{exp}=837\,\mathrm{ks}$, it is the second brightest magnetar \cite[unabsorbed flux $\approx 2.4\times10^{-11}\,\mathrm{erg\,cm^{-2}\,s^{-1}}$, see][]{rea+07b,ok14}, located in Scorpio (R.A. $17^\mathrm{h}$ $08^\mathrm{m}$ $46^\mathrm{s}.3$, DEC. $-40^\circ$ $08'$ $44''.6$) at an estimated distance between $5$ and $10\,\mathrm{kpc}$ \cite{isr+99}. 

In agreement with previous observations \cite{isr+99,rea+07b}, the pulse profile detected by {\it IXPE} is essentially single-peaked, with only a small secondary peak,  much lower than the primary one. The timing analysis allowed to derive a spin frequency $f=0.090795742(5)\,\mathrm{Hz}$ and a frequency derivative $\dot{f}=-1.87(25)\times10^{-13}\,\mathrm{Hz}\,\mathrm{s}^{-1}$ (MJD epoch 59850.84175), which are in-line with previous estimates \cite{dk14} and provide a dipolar magnetic field strength of $B\approx5\times10^{14}\,\mathrm{G}$.

\begin{figure}[h]
\begin{center}
\includegraphics[width=10.7cm]{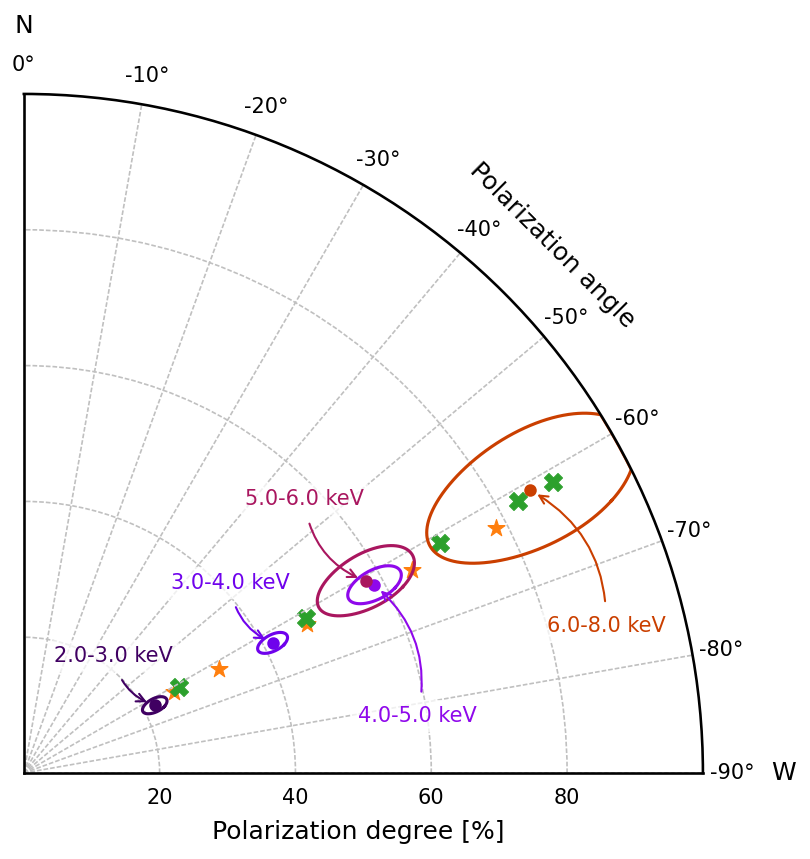}
\caption{Polar plot showing the phase-integrated, energy-dependent PD and PA (filled circles with $1\sigma$ confidence contours) measured by {\it IXPE} for the AXP 1RXS J1708. The results of numerical simulations obtained for model A (orange stars) and model B (green crosses) are also shown (see text for details). Figure taken from \cite{zane+23}. \label{fig:endeppol1708}}    
\end{center}
\end{figure}

\begin{figure*}[h]
\begin{adjustwidth}{-\extralength}{0cm}
\centering
\includegraphics[width=16.7cm]{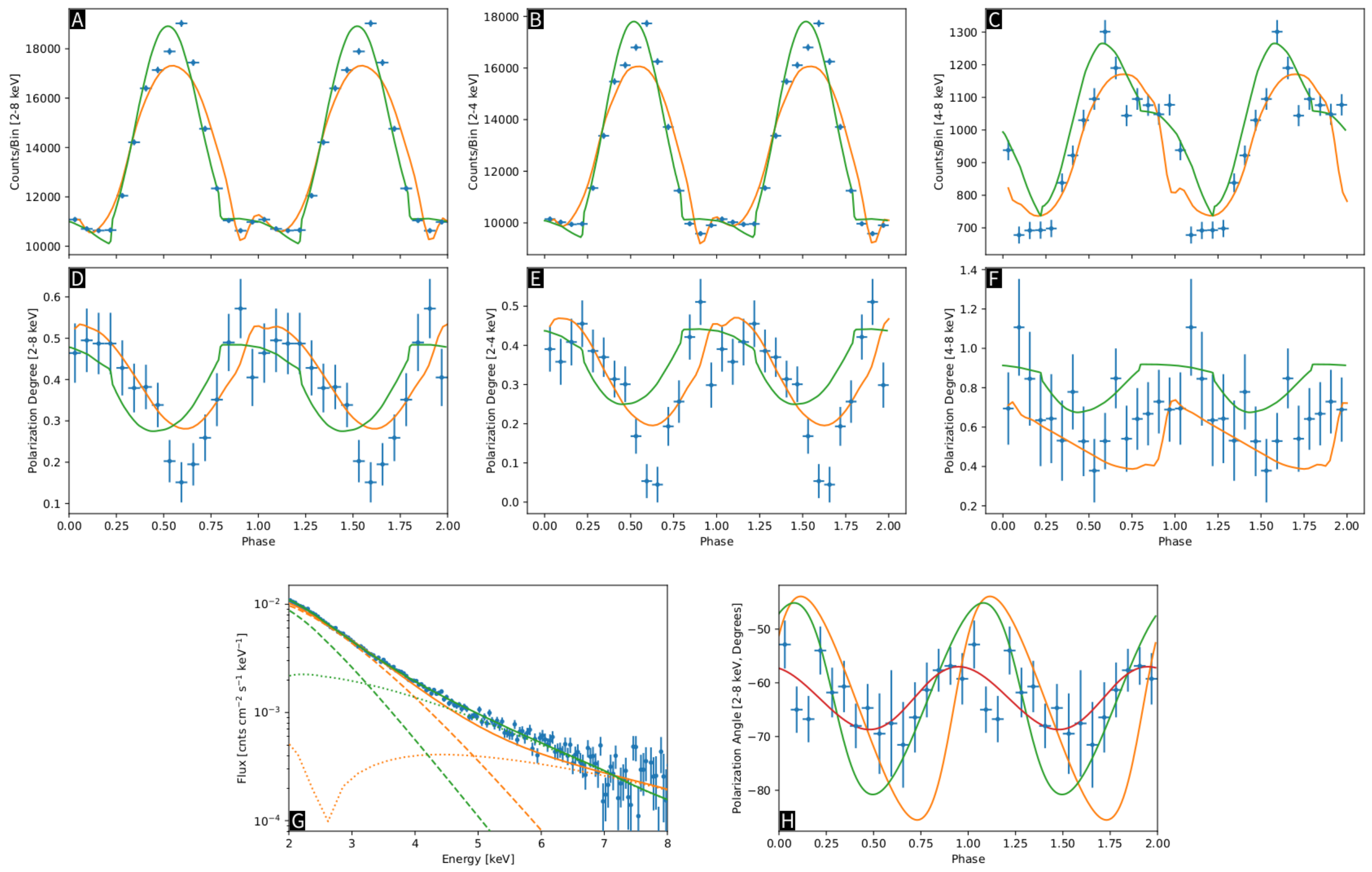}
\caption{Top: {\it IXPE} measurement (cyan filled circles with error bars) of the phase-dependent flux from the AXP 1RXS J1708, energy-integrated in the $2$--$8$ (A), $2$--$4$ (B) and $4$--$8\,\mathrm{keV}$ (C) energy ranges. Center: same for the phase-dependent polarization degree (D to F). Bottom: phase-averaged $2$--$8\,\mathrm{keV}$ spectrum (G) and phase-dependent PA (H) integrated in the entire {\it IXPE} band. Orange (green) solid curves represent the results of numerical simulations performed for model A (model B). The dashed and dotted lines in panel G refer to the spectra of the single emitting regions in models A and B (see text for more details). The red solid line in panel H represents the best fitting RVM. Figure taken from \cite{zane+23}. \label{fig:phdeppol1708}}
\end{adjustwidth}
\end{figure*}

A detailed analysis of the {\it IXPE} polarization measurement (summed over the three DUs) was performed by \citet{zane+23}, after background and spurious modulation subtraction \cite[see][]{dimar+23}. A $\mathrm{PD}=35\pm1.6\%$ was measured (with a significance $\approx 22.5\sigma$), much higher than the $\mathrm{MDP}_\mathrm{99}$ in the same range ($\approx 5\%$). The polarization direction turned out to be $62.1^\circ\pm1.3^\circ$ measured West of North. Following the same procedure already illustrated in \S\ref{subsec:4u} for 4U 0142+61, the {\it IXPE} band was divided into six energy intervals to study the polarization evolution as a function of the photon energy (see Figure \ref{fig:endeppol1708}). At variance with the first source, the polarization angle is quite constant with the energy, at about $60^\circ$ West of North in each bin. The trend of the polarization degree is different as well: it monotonically increases from $21.7\pm1.7\%$ ($\mathrm{MDP}_{99}\approx7\%$) at $2$--$3\,\mathrm{keV}$ to $85\pm15\%$ ($\mathrm{MDP}_{99}\approx50\%$) at $6$--$8\,\mathrm{keV}$. The results of the phase-dependent polarimetric analysis, with data folded at the timing solution mentioned above and divided into 16 phase intervals, are shown in Figure \ref{fig:phdeppol1708}, where the phase-resolved flux and PD are reported at low ($2$--$4\,\mathrm{keV}$) and high ($4$--$8\,\mathrm{keV}$) energies, as well as in the entire {\it IXPE} band. At higher energies the pulse profile turns out to be substantially different from those at low energies and integrated over the entire $2$--$8\,\mathrm{keV}$ band: the decrease of counts in the primary peak makes, in fact, the overall profile double-peaked. The phase-dependent profile of the polarization degree is in counterphase with respect to the light curve. 
However, the low number of counts collected between $4$ and $8\,\mathrm{keV}$ makes the high-energy polarization trend rather uncertain. For what concerns the polarization direction, also in this case the phase-dependent PA behavior is uncorrelated with the polarization degree, exhibiting an essentially sinusoidal trend, much in the same way as observed for 4U 0142+61.

In order to perform a spectro-polarimetric analysis, 
the count statistics were increased by exploiting a quasi-si\-mul\-tan\-eous {\it Swift-XRT} \cite{bur+05} observation, with an exposure of $\approx 1\,\mathrm{ks}$, performed on October 20, 2022, and also a close observation by {\it NICER} \cite{arz+14}, performed on August 21, 2022. The phase-integrated Stokes parameter data were fitted simultaneously inside XSPEC, using either a (absorbed) BB+BB or BB+PL decomposition and convolving each spectral component with a constant polarization model ({\tt polconst} in XSPEC). Both the fits turned out to be statistically acceptable \cite{zane+23} and, 
for the BB+PL spectral decomposition, the fit parameters were not much different with respect to those reported by previous observations, 
although the  column density ($N_\mathrm{H}=2.09^{+0.06}_{-0.05}\times10^{22}\,\mathrm{cm}^{-2}$) exceeds that reported in \cite{rea+07b} by more than $3\sigma$. 
Both the BB and the PL turned out to be highly polarized, with polarization directions set apart by $90^\circ$. Even if statistically acceptable, this scenario is not in agreement with the twisted magnetosphere model (see \S\ref{subsec:RCS}), according to which PD should be not much higher than $\approx 30\%$ at higher energies where upscattered photons dominate the spectrum. On the other hand, a purely thermal (BB+BB) decomposition returns a column density $N_\mathrm{H}=1.39^{+0.04}_{-0.04}\times10^{22}\,\mathrm{cm}^{-2}$ and a colder BB temperature $kT_\mathrm{c}=0.44^{+0.01}_{-0.01}\,\mathrm{keV}$, both in excellent agreement with the values derived from previous analyses. Moreover, with a hotter BB temperature $kT_\mathrm{h}=1.07\pm0.03\,\mathrm{keV}$, the two components are well separated across the {\it IXPE} band, so that the polarization parameters can be more easily reconciled with the polarimetric results discussed above \cite[i.e. a lower (higher) PD for the colder (hotter) component, with basically the same polarization orientation, see][]{zane+23}.

\citet{zane+23} proposed a model for the emission from 1RXS J1708 in terms of two distinct emitting regions on the NS surface: the first one covered by a magnetized atmospheric layer (to explain the high degree of polarization detected at high energies, see \S\ref{subsec:atmo}), and the second one with the solid surface exposed (so as to account for the lower polarization degree at low energies). This scenario is consistent with the BB+BB decomposition obtained from the spectral analysis. Since the surface temperature of a NS is fixed by the inclination of the $B$-field wrt the local normal \cite[e.g.][]{ppp15}, it is plausible that different surface regions have different temperatures and, in particular, that the hotter part is covered by a gaseous layer, which becomes a magnetic condensate where the temperature has decayed enough. The simulation results shown in Figures \ref{fig:endeppol1708} and \ref{fig:phdeppol1708} have been obtained using two different flavours of this scenario: model A, in which the hotter, atmospheric region is a circular hot-spot located at one of the magnetic poles, while the colder, condensed-surface region is an equatorial belt (similar to that of 4U 0142+61, see \S\ref{subsec:4u}); and model B, where the two different regions are spots with different size, located at different positions on the star surface. In both models A and B, the atmospheric emission properties have been calculated using the code described in \cite{lloyd03} (without mode conversion effects), while the condensed surface emission has been modelled according to Potekhin et al. \cite{pot+12}. The two simulations can  successfully reproduce the increase of PD with the photon energy at a constant PA, as well as the phase-dependent behavior of the polarization degree in the different energy bands. The best agreement is found for $\chi\approx30^\circ$ and $\xi\approx10^\circ$ for both models. Such values turn out to be compatible with those derived by interpreting the hard X-ray spectrum of 1RXS J1708 in terms of the coronal outflow model by \citet[see \citenum{hbh14}]{belo13}. 

The observed anti-correlation in the pulse profiles of the flux and the polarization degree can be explained if the condensed surface region, which is responsible for the less polarized flux, is more extended than the atmospheric cap, which in turn emits highly polarized radiation. In both model A and B, this conditions is met (the ratio of the areas is $\approx 5\%$) and the temperatures of the two zones are close ($\approx 0.7\,\mathrm{keV}$). The minimum of the flux corresponds to a configuration in which photons coming from the condensed region do not reach the observer, so that those from the atmospheric cap contribute most. On the other hand, at the pulse maximum both the regions are visible (at least for the viewing geometry we inferred), but the flux is dominated by the less polarized radiation from the condensate.
Finally, despite the complicated shape of the PD pulse profiles, both model A and B predict a sinusoidal shape for the phase-resolved polarization angle, in agreement with observations.

\subsection{SGR 1806--20} \label{subsec:1806}
The SGR 1806--20 is mostly known for the emission of the most energetic giant flare ever recorded from a magnetar,  on December 27, 2004 \cite{palm+05}. Its position in the sky is close to the Galactic center (R.A. $18^\mathrm{h}$ $08^\mathrm{m}$ $39^\mathrm{s}.8$, DEC. $-20^\circ$ $24'$ $26''.7$), at an estimated distance $\approx 8.7\,\mathrm{kpc}$ \cite{bibb+08}. The source was observed by {\it IXPE} between March 22 and April 13, 2023, for a total exposure time $t_\mathrm{exp}\approx947\,\mathrm{ks}$, and a simultaneous observation with {\it XMM-Newton} \cite{jans+01} was carried out on April 7, 2023, with $t_\mathrm{exp}\approx45\,\mathrm{ks}$ \cite{tur+23}. Unfortunately the record-low flux ($F_\mathrm{unabs}\approx 4\times10^{-12}\,\mathrm{erg}\,\mathrm{cm}^{-2}\,\mathrm{s}^{-1}$, less than half of the historical value, $\approx2\times10^{-11}\,\mathrm{erg}\,\mathrm{cm}^{-2}\,\mathrm{s}^{-1}$\cite[][]{ok14}) and the occurrence of a large number of solar flares that contaminated the {\it IXPE} observation made the counting statistics too low for a satisfactory analysis.

The timing analysis based on the {\it IXPE} data did not give significant solutions. A spin frequency of $f=0.128695(3)\,\mathrm{Hz}$ was instead extracted from the  {\it XMM-Newton} data (MJD 57202.0) and is compatible with the most recent estimate by \cite{youn+17}. Much in the same way, spectral fits to the  {\it IXPE} counts did not allow to reach a definite conclusion, while both a BB+BB and a BB+PL model provide an acceptable fit to  {\it XMM-Newton} EPIC data. The inferred parameters for the latter model are compatible with those obtained  over the last 10 years, even if the source flux is remarkably lower (by nearly  a factor of 3).

\begin{figure}[h]
\begin{center}
\includegraphics[width=9.7cm]{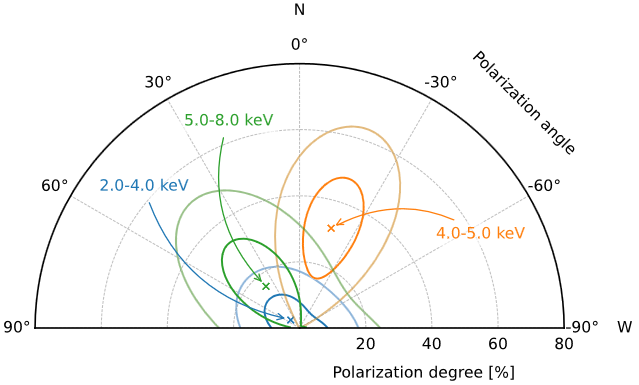}
\caption{Polar plot showing the phase-integrated, energy-dependent PD and PA (crosses with  $68.3\%$ and $99\%$ confidence contours) measured by {\it IXPE} for SGR 1806--20. Figure taken from \cite{tur+23}. \label{fig:endeppol1806}}    
\end{center}
\end{figure}

Only about $8000$ background-subtracted events were collected in the three DUs and, as a consequence, the phase- and
energy-integrated ($2$–$8\,\mathrm{keV}$) polarization degree ($5.7\%$) is well below the $\mathrm{MDP}_{99}\approx 20\%$ and is not significant; the polarization angle is not constrained.
By performing an energy-resolved,
phase-integrated analysis, however, a signal was found in the $4$--$5\,\mathrm{keV}$ bin with $\mathrm{PD} = 31.6\% \pm 10.5\%$ (marginally above the $\mathrm{MDP}_{99}$), and $\mathrm{PA}={17.6^\circ}^{+15.5^\circ}_ {-15.0^\circ}$ computed East
of North. Only an upper limit of $24\%$ and $55\%$ ($99\%$ confidence level) was found  in the two
neighboring bins, $2$--$4\,\mathrm{keV}$ and $5$--$8\,\mathrm{keV}$, respectively (see Figure \ref{fig:endeppol1806}). 

The  upper limits found at low and high
energies, together with the $99\%$ confidence level measurement of $\mathrm{PD}\approx 32\%$ at $4$--$5\,\mathrm{keV}$, are compatible with a scenario where thermal photons coming from a region of the condensed star's surface are then reprocessed by magnetospheric RCS. The source, then, may be similar to 4U 0142+61 (see \S\ref{subsec:4u}), although present data do not allow to claim that low energy photons are mostly polarized in the O-mode and high energy ones in the X-mode, since no $90^\circ$ swing in PA is presently seen. 

\subsection{AXP 1E 2259+586} \label{subsec:2259}
The last magnetar observed by {\it IXPE} so far, the AXP 1E 2259+586, is located in Cassiopeia (R.A. $23^\mathrm{h}$ $01^\mathrm{m}$ $08^\mathrm{s}.8$, DEC. $58^\circ$ $52'$ $20''.8$), at a distance $3.2\,\mathrm{kpc}$ \cite{kf12}. It was observed between June 2 and July 6, 2023, with a net exposure time $t_\mathrm{exp}=1.2\,\mathrm{Ms}$. Simultaneous data were taken with {\it XMM-Newton} (on June 30, 2023, for $\approx 19\,\mathrm{ks}$) and {\it NICER} \cite[as a part of a wider campaign from March 19 and July 24, 2023, see][]{heyl+23}.

The spin frequency obtained from a timing analysis of the joint {\it IXPE}, {\it XMM-Newton} and {\it NICER} data turned out to be $f=0.143281286(2)\,\mathrm{Hz}$ (MJD 60022.0). Moreover, thanks to the particularly long {\it NICER} observation, an accurate estimate of the frequency derivative was possible, with $\dot{f}=-9.7(3)\times10^{-15}\,\mathrm{Hz}$. These values are in excellent agreement with previous estimates \cite{dk14}, implying a dipolar component of the magnetic field with strength $B\approx6\times10^{13}\,\mathrm{G}$.

\begin{figure}[H]
\begin{center}
\includegraphics[width=8.0cm]{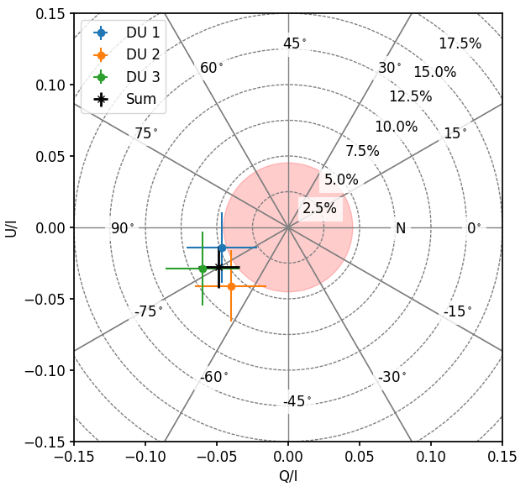}
\caption{Normalized Stokes parameters $Q/I$ and $U/I$ (filled circles with error bars) measured for the AXP 1E 2259+586 by DU 1 (cyan), 2 (orange) and 3 (green). The correspondent measurement obtained by summing together the contributions of the three DUs is marked by the black cross with $1\sigma$ error bars. The gray dashed curves and solid lines represent the loci of constant PD and PA, respectively (the corresponding values are reported in the plot). The red-shaded area indicates the $\mathrm{MDP}_{99}$ in the $2$--$8\,\mathrm{keV}$ range. Figure taken from \cite{heyl+23}. \label{fig:phenintpol2259}}    
\end{center}
\end{figure}

Spectral fits of the {\it XMM-Newton} data in the $0.5$--$8\,\mathrm{keV}$ hint for the presence of a PL component, as already pointed out in literature \cite{zhu+08}. Although the best fitting parameters are in agreement with those listed in the McGill catalogue \cite{ok14}, a BB+PL spectral decomposition is not entirely satisfactory (reduced $\chi^2$ $\approx 2$ for $248$ dof). Indeed, the fit to both EPIC-pn and {\it IXPE} data improved significantly by adding an absorption line, as already found 
in the most recent spectral observation of AXP 1E 2259+586 by \citet{pizz+19}; the  line energy ($0.96^{+0.07}_{-0.18}\,\mathrm{keV}$) and width ($0.23^{+0.10}_{-0.06}\,\mathrm{keV}$) are in good agreement, within the errors, with those reported in \cite{pizz+19}.

\begin{figure}[H]
\begin{center}
\includegraphics[width=9.7cm]{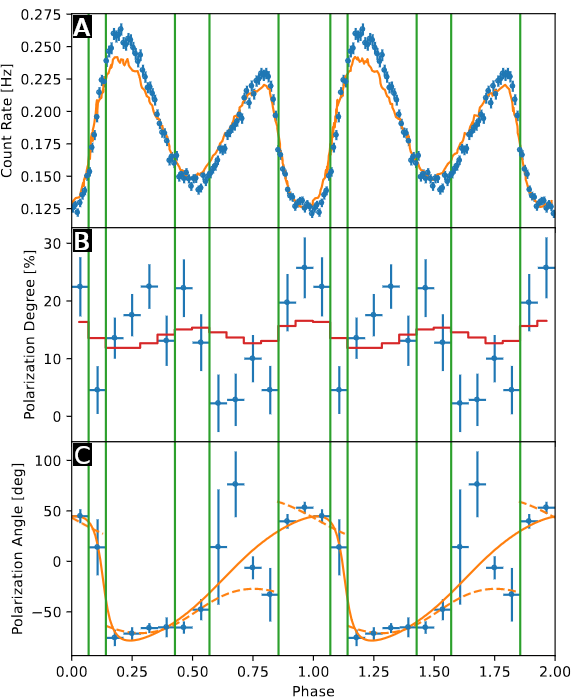}
\caption{{\it IXPE} measured counts (panel A), polarization degree (panel B) and polarization angle (panel C) as a function of the rotational phase, integrated in the $2$--$8\,\mathrm{keV}$ band (filled circles). The reported error bars correspond to $\Delta\log L=1/2$ contours of the likelihood $L$ \cite{gonz+23}. The orange solid curve in panel A represents the $0.3$--$12\,\mathrm{keV}$ light curve observed by {\it XMM-Newton} in 2014 \cite{pizz+19}. The red solid curve in panel B denotes the $\mathrm{MDP}_{99}$. The orange solid and dashed curves in panel C correspond to the best fitting RVM model in the hypothesis of a fixed polarization mode and two different normal modes, respectively. The green vertical lines highlight the relevant intervals of the pulse profile: primary dip, rise, primary peak, secondary dip and secondary peak. Figure taken from \cite{heyl+23}. \label{fig:phdeppol2259}}    
\end{center}
\end{figure}

The phase-integrated polarization measurement over the entire {\it IXPE} band revealed polarized emission above the $\mathrm{MDP}_{99}$ (which amounts to $4.5\%$). Joining together the counts of the three detectors gives $\mathrm{PD}=5.6\pm1.4\%$ and $\mathrm{PA}=-75^\circ.2\pm7^\circ.4$ (measured East of North, see Figure \ref{fig:phenintpol2259}). As a function of the photon energy, the polarization turned out to be significant only at low energies ($2$--$3\,\mathrm{keV}$), with $\mathrm{PD}=6.1\pm1.5\%$ ($\mathrm{MDP}_{99}=4.6\%$) and $\mathrm{PA}=66^\circ.4\pm7^\circ.1$ West of North. At higher energies only upper limits (at $3\sigma$ confidence level) are available, with $\mathrm{PD}<14.6\%$ at $3$--$5\,\mathrm{keV}$ and $<70.0\%$ at $5$--$8\,\mathrm{keV}$. Interestingly, the phase-dependent behavior of the polarization properties (integrated over the entire {\it IXPE} band) is quite complex, following essentially the double-peaked light-curve, as shown in panels A--B of Figure \ref{fig:phdeppol2259}). A high polarization degree $\approx20$--$26\%$ ($\mathrm{MDP}_{99}\approx16$--$17\%$) is detected in correspondence to the primary dip of the pulse profile. It then vanishes as the flux increases towards the primary peak ($\mathrm{PD}\approx5\%$, below the $\mathrm{MDP}_{99}\approx 14\%$), to rise again to $\approx 23\%$ ($\mathrm{MDP}_{99}=13\%$) in the secondary dip. Finally, the polarization returns to be compatible with zero (i.e. below the $\mathrm{MDP}_{99}$) in the subsequent rise and secondary peak. Correspondingly the polarization direction swings between $\approx 45^\circ$ and $\approx -75^\circ$ (measured East of North) assuming a sinusoidal trend.

\citet{pizz+19} suggested that a plasma loop close to the star's surface can act as a sort of ``screen'', (partly) intercepting the radiation from the surface at certain rotational phases. 
This is similar to the scenario proposed by \citet{tiengo+13} for another magnetar candidate, the transient source SGR 0418+5729, in which a phase-variable absorption line was detected. \citet{heyl+23} related the peculiar behavior of PD and PA in the AXP 1E 2259+596 to the effects of such a magnetic loop. The primary dip in the light curve would correspond to the case in which  most of the photons, emitted by a hotter region of the surface facing the observer, are almost completely intercepted by the loop and so scattered off the LOS. As the star rotates, the  positions of the loop and the emitting region change, allowing more photons from the surface to reach the observer and producing the rise of the flux towards the primary peak of the pulse. Then, another emitting region should enter into view (as suggested, for example, in the case of the AXP 1RXS J1708, see \S\ref{subsec:1708}), once the radiation from the primary spot (covered by the magnetic loop) is definitely outside the field of view. The observed phase-dependent behavior of the polarization properties would follow if, as hinted in \cite{pizz+19}, the main interaction between photons and plasma particles is through scattering off protons at the cyclotron resonance. The (super-strong) magnetic field in the loop  modifies the scattering cross sections, in such a way that scattering 
into X-mode photons is more likely than that into the O-mode ones (see \S\ref{subsec:RCS}). In the assumption that thermal radiation is only mildly polarized in the O-mode (as e.g. for the AXP 4U 0142+61, see \S\ref{subsec:4u}), radiation in the primary dip of the light curve is mostly O-mode photons, that underwent a small number of interactions.
As the loop moves away from the LOS, photons of the hotter spot, that underwent many scatterings and, therefore, are more likely polarized in the X-mode, are collected. This causes an initial decrease of the polarization degree in the rise phase of the light curve (due to the swing of the polarization direction), followed by a further increase once the primary peak is reached. Then, as the loop and the hotter spot depart from the LOS, photons coming directly from the secondary spot are collected. Since they are not affected by scattering as before, photons of the secondary dip and secondary peak in the pulse profile exhibit only a negligible polarization degree. Actually, as shown in panel C of Figure \ref{fig:phdeppol2259}, also in this case the phase-dependent behavior of the polarization angle can be fitted by the RVM, and the fits are equally acceptable either considering that photons are polarized in the same polarization mode at all phases or assuming a mode switching (as required in the scenario discussed here). However, the probability to explain the entire observation without the mode switching turned out to be smaller than $3\times10^{-4}$ \cite{heyl+23}.

\section{Discussion and Conclusions} \label{sec:concl}

In this work we have summarized the results of the polarization measurements performed by {\it IXPE} in the $2$--$8\,\mathrm{keV}$ band of radiation coming from four magnetar sources. These first X-ray polarization measurements performed more than 40 years after the launch of the {\it OSO-8} satellite, thanks to GPD polarimeters based on the photoelectric effect, have opened a new window in the study of strongly magnetized NSs, complementing the information coming from spectral and timing analyses.

The main findings that we have discussed in this paper can be summarized as follows.
\begin{itemize}
\item As expected for source endowed with ultra-strong magnetic fields, magnetar emission turned out to be strongly polarized. The polarization degree observed in the four objects  ranges from $\approx 15$--$20\%$ at low energies ($2$--$4\,\mathrm{keV}$) to more than $80\%$ in the higher end of the {\it IXPE} band. 
\item Despite the similarities in spectral shape shared by the four magnetars, the different polarization patterns indicate that the thermal emission may have different origins.  It may come directly from regions where the condensed surface is exposed or being reprocessed in a geometrically-thin, magnetized atmosphere above the crust. In this respect, polarimetry can indeed provide a way to disentangle different emission models, removing the degeneracy of spectral analysis alone.
\item The peculiar $90^\circ$ swing of the polarization angle detected in the AXP 4U 0142+61 can be naturally explained in terms of photons polarized in two normal modes, the ordinary and extraordinary ones. Since such pattern is expected if radiation propagates in magnetic fields $B\gtrsim B_\mathrm{Q}$, this can be regarded as an indirect proof that magnetar magnetic fields are indeed ultra-strong. Moreover, the limited value of the polarization degree detected at high energies ($\approx30$--$35\%$) argues in favor of the canonical twisted magnetosphere scenario, according to which resonantly up-scattered photons dominate the high-energy part of the soft X-ray spectrum.
\item Even if the phase- and energy-integrated measurements did not yield a polarization degree high enough to validate the presence of vacuum birefringence \cite[$\gtrsim40\%$, see][]{tav+20}, the detected phase-dependent behavior of the polarization degree and angle (with PD following the flux pulse profile and PA modulated according to the RVM) is indeed what one expects if vacuum birefringence is at work. This can be considered as a first step towards testing QED effects in the ultra-magnetized vacuum.
\item Phase-resolved polarimetry can be as well a powerful tool in understanding the star magnetic field topology. As in the case of the AXP 1E 2259+586, if a sufficiently high number of photons is collected, the variation of the polarization degree and angle with rotational phase can help in confirming the existence of plasma loops, responsible for the occurrence of phase-dependent absorption lines in the spectrum. In this respect, an important contribution in improving the spectro-polarimetric, phase-dependent analysis may come from forthcoming X-ray missions, like HEX-p \cite[][]{alf+24} and eXTP \cite[][]{zhang+16}.
\end{itemize}
The X-ray polarization measurements by the {\it IXPE} observatory have demonstrated that polarimetry is crucial to foster our understanding of ultra-magnetized neutron stars. Looking at the future, it will be important to extend polarmetric measurements towards lower energies ($0.1$--$2\,\mathrm{keV}$), where magnetar radiation is peaked. Nevertheless, further observations with {\it IXPE} will improve present measurements, especially if  transient magnetars in outburst are among the targets since this can help in providing further evidence of vacuum birefringence effects.



\vspace{6pt}

\funding{This research was funded by the Italian Ministry of University and Research (MUR) through grant PRIN 2022LWPEXW.}

\dataavailability{{\it IXPE} data and response functions are available in the HEASARC online data archive \url{https://heasarc.gsfc.nasa.gov/docs/ixpe/archive/}.} 




\conflictsofinterest{The authors declare no conflicts of interest.} 

\begin{adjustwidth}{-\extralength}{0cm}

\reftitle{References}

\PublishersNote{}
\end{adjustwidth}
\end{document}